\documentclass[sigconf]{acmart}
\AtBeginDocument{%
  }
\usepackage{enumitem}

\newcommand{\etal}{{\emph{et al.}\ }}

\newcommand{\ie}{{\emph{i.e.,}\ }}

\newcommand{\name}[0]{BISTRO}

\newtheorem{defi}{Definition}

\newtheorem{pro}{Problem}

\usepackage{array}
\usepackage{booktabs}
\usepackage{csvsimple}
\usepackage{multirow}
\usepackage{balance}

\makeatother 

\usepackage{tikz}
\usepackage{pgfplots}
\pgfplotsset{width=8cm,compat=1.13}
\usepackage{arydshln} % dashed line
\usetikzlibrary{arrows,shapes,shadows,snakes,positioning,automata,backgrounds,patterns}
\usepackage{graphicx}
\usepackage{subfigure}
\usepackage{caption}
\usepackage{subcaption}
\usepackage{algorithm, algorithmic}

\usepackage{enumitem}
\usepackage{amsmath}

\usepackage{caption}
\usepackage{xcolor}
\usepackage{color}

\definecolor{mycolor51}{RGB}{84,89,105}    % 1 black
\definecolor{mycolor52}{RGB}{164,117,125}  % 2 brown
\definecolor{mycolor53}{RGB}{231,152,124} % 3 orange
\definecolor{mycolor54}{RGB}{139,145,182}   % 4 blue
\definecolor{mycolor55}{RGB}{119,113,164} % 5 purple

\definecolor{mycolor55light}{RGB}{171,164,195}

\definecolor{mycolor41}{RGB}{54,80,131}
\definecolor{mycolor42}{RGB}{183,131,175}
\definecolor{mycolor43}{RGB}{245,166,115}
\definecolor{mycolor44}{RGB}{252,219,114}

\definecolor{mycolor42light}{RGB}{206,178,203}
\definecolor{mycolor42light2}{RGB}{222,204,221}
\definecolor{mycolor43light}{RGB}{245,201,169}
\definecolor{mycolor44light}{RGB}{249,230,175}

\definecolor{mycolor61}{RGB}{17,50,93}
\definecolor{mycolor62}{RGB}{54,80,131}
\definecolor{mycolor63}{RGB}{115,107,157}
\definecolor{mycolor64}{RGB}{183,131,175}
\definecolor{mycolor65}{RGB}{245,166,115}
\definecolor{mycolor66}{RGB}{252,219,114}

\definecolor{mycolor31}{RGB}{189,142,192}
\definecolor{mycolor32}{RGB}{111,128,190}
\definecolor{mycolor33}{RGB}{244,126,98}

\copyrightyear{2024}
\acmYear{2024}
\setcopyright{acmlicensed}
\acmConference[KDD '24]{Proceedings of the 30th ACM SIGKDD Conference on Knowledge Discovery and Data Mining}{August 25--29, 2024}{Barcelona, Spain}
\acmBooktitle{Proceedings of the 30th ACM SIGKDD Conference on Knowledge Discovery and Data Mining (KDD '24), August 25--29, 2024, Barcelona, Spain}
\acmDOI{10.1145/3637528.3671759}
\acmISBN{979-8-4007-0490-1/24/08}
\acmPrice{}

\begin{document}

% 调整间距
% \setlength{\intextsep}{9pt} % 调整图形与周围文本之间的间距
% \setlength{\abovecaptionskip}{5pt} % 调整标题上方的间距
% \setlength{\belowcaptionskip}{0pt}  % 调整标题下方的间距

\title{Adapting Job Recommendations to User Preference Drift with Behavioral-Semantic Fusion Learning}


\author{Xiao Han}
\email{hahahenha@gmail.com}
% \authornotemark[2]
\orcid{0000-0002-3478-964X}
\affiliation{%
  \institution{City University of Hong Kong}
  % \streetaddress{83 Tat Chee Avenue}
  \city{Hong Kong}
  \country{China}
  % \postcode{999077}
}

\author{Chen Zhu}
\email{zc3930155@gmail.com}
\authornotemark[1]
% \orcid{0000-000x-xxxx-xxxx}
\affiliation{%
  \institution{Career Science Lab, BOSS Zhipin \\ University of Science and Technology of China}
  % \streetaddress{xxxx}
  \city{Beijing}
  % \state{Beijing}
  \country{China}
  % \postcode{xxxxxx}
}

\author{Xiao Hu}
\email{huxiao@kanzhun.com}
% \orcid{0000-000x-xxxx-xxxx}
\affiliation{%
  \institution{Career Science Lab, BOSS Zhipin}
  % \streetaddress{xxxx}
  \city{Beijing}
  % \state{Beijing}
  \country{China}
  % \postcode{xxxxxx}
}

\author{Chuan Qin}
\email{chuanqin0426@gmail.com}
% \orcid{0000-000x-xxxx-xxxx}
\affiliation{%
  \institution{Career Science Lab, BOSS Zhipin}
  % \streetaddress{xxxx}
  \city{Beijing}
  % \state{Beijing}
  \country{China}
  % \postcode{xxxxxx}
}

\author{Xiangyu Zhao}
\email{xianzhao@cityu.edu.hk}
\authornotemark[1]
\affiliation{
  \institution{City University of Hong Kong}
  % \streetaddress{83 Tat Chee Avenue}
  \city{Hong Kong}
  \country{China}
  % \postcode{999077}
}

\author{Hengshu Zhu}
\email{zhuhengshu@kanzhun.com}
\authornotemark[1]
% \orcid{0000-000x-xxxx-xxxx}
\affiliation{%
  \institution{Career Science Lab, BOSS Zhipin}
  % \streetaddress{xxxx}
  \city{Beijing}
  % \state{}
  \country{China}
  % \postcode{xxxxxx}
}

\thanks{* Chen Zhu, Xiangyu Zhao, and Hengshu Zhu are the
corresponding authors.}
\thanks{\dag
This work was accomplished by the first author while interning at BOSS Zhipin under the supervision of the second author.}

\renewcommand{\shortauthors}{Xiao Han et al.}


% a session-based framework, \name, is developed, aiming to accurately reflect users' real-time job preferences through behavioral-semantic fusion learning.

\begin{abstract}

Job recommender systems are crucial for aligning job opportunities with job-seekers in online job-seeking. However, users tend to adjust their job preferences to secure employment opportunities continually, which limits the performance of job recommendations. 
The inherent frequency of preference drift poses a challenge to promptly and precisely capture user preferences. To address this issue, we propose a novel session-based framework, \name, to timely model user preference through fusion learning of semantic and behavioral information. Specifically, \name\ is composed of three stages: 1) coarse-grained semantic clustering, 2) fine-grained job preference extraction, and 3) personalized top-$k$ job recommendation. Initially, \name~segments the user interaction sequence into sessions and leverages session-based semantic clustering to achieve broad identification of person-job matching. Subsequently, we design a hypergraph wavelet learning method to capture the nuanced job preference drift. To mitigate the effect of noise in interactions caused by frequent preference drift, we innovatively propose an adaptive wavelet filtering technique to remove noisy interaction. Finally, a recurrent neural network is utilized to analyze session-based interaction for inferring personalized preferences. Extensive experiments on three real-world offline recruitment datasets demonstrate the significant performances of our framework. Significantly, \name\ also excels in online experiments, affirming its effectiveness in live recruitment settings. This dual success underscores the robustness and adaptability of \name.

\end{abstract}




\received{8 February 2024}
\received[revised]{5 April 2024}
\received[accepted]{16 May 2024}

\maketitle

\section{Introduction}

\begin{figure}[htb!]
\centering
\subfigure[A scenario illustrating a user revise the resume to target a different job position]{
\makebox[\linewidth][c]{
\includegraphics[width=\linewidth]{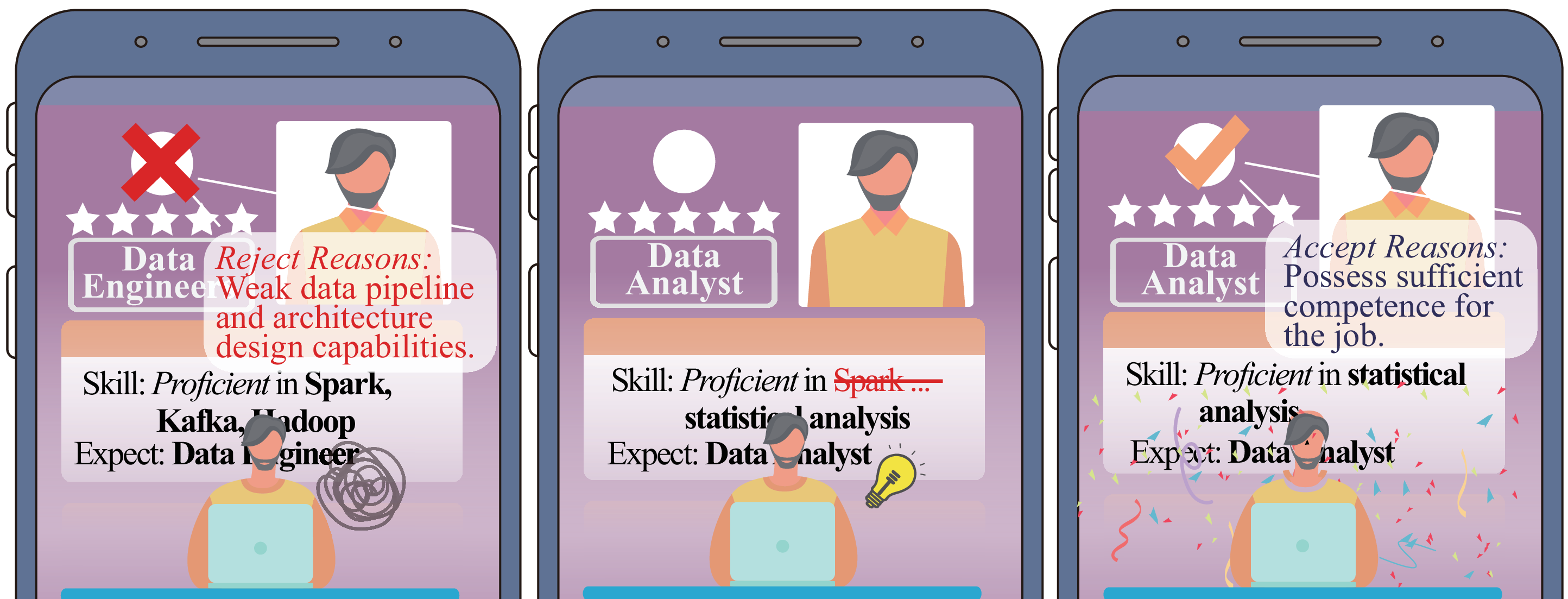}}
\label{fig:1a}
}

\subfigure[Statistics on revision of resumes]{
\makebox[0.3\linewidth][c]{
\includegraphics[width=0.3\linewidth]{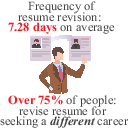}}
\makebox[0.55\linewidth][c]{
\includegraphics[width=0.55\linewidth]{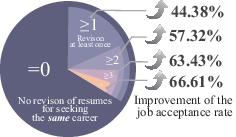}}
\label{fig:1b}
}
\caption{A toy example in the online recruitment platform.}
\label{fig:1}
% \vspace{-5mm}
\end{figure}

% \begin{figure}[htb!]
% \centering
% \makebox[\linewidth][c]{
% \includegraphics[width=0.6\linewidth]{fig/fig1b.eps}}
% \caption{The scenario shows a user could modify resumes to meet the job requirements.
% \re{Need to revise (two figures, 1) change job-seeking intents when doing job-searching; 2) statics of sessions, resume revised, \etc)}
% }
% \label{fig:1}
% % \vspace{-5mm}
% \end{figure}

In recent years, a clear trend has emerged where online recruitment platforms are undergoing rapid development and surpassing local job markets as the primary recruitment channel.
According to the market research report \cite{ortmfr2030}, the global online recruitment market size, which was valued at USD 29.29 billion in 2021, is projected to expand to USD 58.16 billion by 2030, with a compound annual growth rate of 7.1\% from 2023 to 2030.
Consequently, the job recommender system, a primary method in online recruitment, has also underscored its significance.

However, unlike conventional recommendation settings, recruitment is inherently a bidirectional selection process \cite{DBLP:conf/aaai/LinZZZWX17,DBLP:conf/aaai/WangZZQX20}.
This dynamic necessitates that not only should the jobs align with the expectations of the job-seekers, but the job-seekers must also satisfy the requirements set by the employers.
As a toy example illustrated in Figure~\ref{fig:1a}, active users refine job preferences in order to secure employment opportunities, reflected in the user-job interactions and the resume revision.
Initially, the individual seeks employment as a data engineer but encounters setbacks due to the shortage of data pipeline and architecture design capabilities.
This experience led to the realization that his skills may be more aligned with data analyst positions, which shows the preference drift.
Motivated by this insight, the job-seeker refines his resume to better match data analyst positions, culminating in the successful acquisition of an offer.
% However, unlike conventional recommendation settings, recruitment is inherently a bidirectional selection process.
% This dynamic necessitates that not only should the jobs align with the expectations of the job-seekers, but the job-seekers must also satisfy the requirements set by the employers.
% Therefore, during the job-seeking process, job-seekers often iteratively refine their job preferences in order to secure employment opportunities, reflected in user behavior and resume refinement.
% We show a toy example in Figure~\ref{fig:1a} that illustrates a job-seeker's journey.
% Initially, the individual seeks employment as a data engineer but encounters setbacks due to the shortage of data pipeline and architecture design capabilities.
% This experience led to the realization that his skills may be more aligned with data analyst positions.
% Motivated by this insight, the job-seeker refines his resume to better match data analyst positions, culminating in the successful acquisition of an offer.
Furthermore, Figure~\ref{fig:1b} provides some statistical insights about resume update behaviors collected from a prominent online recruitment platform in China over a six-month period to imply the prevalence of this phenomenon.
It highlights that, on average, job-seekers tend to revise their resumes every 7.28 days when they do not secure employment and those who update their resumes frequently are at least 44\% more successful in receiving job offers than the rest of those who are not proactive, underscoring the effectiveness of consistently optimizing resumes during the job-seeking process.
Another interesting insight is over three-quarters of individuals would change their job-seeking objectives when refining their resumes, indicating a strong correlation between preference drift and resume refinement.

The inherent propensity for frequent preference drift implies the necessity of nuanced and timely modeling of user preference in job recommendation, which limits the effectiveness of both content and behavior-based recommendation algorithms.
On the one hand, content-based job recommender systems~\cite{yang2017combining,yadalam2020career,liao2021investigating} strive to precisely profile job-seeker capabilities for better aligning their qualifications with recommendation results. However, they fall short in capturing job-seeker nuanced job preferences. 
On the other hand, although behavior-based recommendation, especially session-aware recommendation~\cite{zhang2020personalized,pang2022heterogeneous,liu2023multi}, provides a solution to timely track job-seeker preferences by short-term interactions, their performance is limited by the length of session-based interactions and thus is vulnerable to noise, which is common in practical job recommender systems.

To address the above issues, we propose a novel session-based framework in this paper, namely \textbf{\name}~(\underline{\textbf{B}}ehav\underline{\textbf{I}}oral-\underline{\textbf{S}}eman\underline{\textbf{T}}ic fusion for job \underline{\textbf{R}}ec\underline{\textbf{O}}mmendation).
Specifically, the framework contains three modules:
1) A coarse-grained semantic clustering module. It groups users or jobs based on semantics to facilitate the broad identification of person-job matching.
2) A fine-grained job preference extraction module. In this module, a multi-granular interaction hypergraph is constructed for capturing preference drift and a novel spectral hypergraph wavelet learning method is performed on this graph to capture preference drift and denoising preference feature.
3) A personalized top-$k$ job recommendation module. It utilizes a recurrent neural network to analyze short-term sequential behavior and infer personalized preferences, thereby generating recommendations for the top-$k$ jobs.
In summary, our contributions are demonstrated as follows:

% To address above issues, in this paper, we propose a novel session-based framework namely \name~(BehavIoral-SemanTic fusion for job RecOmmendation).
% % , where user interaction sequence is first segmented into sessions based on the timestamp of resume updates and then a combination of session-based clustering and hypergraph neural network-based multi-granular user interaction modeling is employed for noise-robust and timely job preference modeling.
% Specifically, the framework contains three modules:
% 1) coarse-grained semantic clustering modules, which 
% 2) fine-grained job preference extraction, and 
% 3) personalized top-$k$ job recommendation.
% Initially, the coarse-grained semantic clustering module groups users (jobs) based on semantics to facilitate the broad identification of person-job matching.
% Subsequently, to capture preference drift, we construct a multi-granular interaction hypergraph~(\ie intra-session and inter-session) and employ a spectral graph network to track the job preferences of users.
% To mitigate the effect of noise, we innovatively propose an adaptive wavelet filtering technique in this stage.
% In the final stage, a recurrent neural network is utilized to analyze short-term sequential behavior and infer personalized preferences, thereby generating recommendations for the top-$k$ jobs.
% In summary, our contributions are demonstrated as follows:

\begin{itemize}[leftmargin=*]
    \item To capture the job preference drift in job-seeking, we innovatively propose a session-based recommendation framework, \name, which leverages behavioral-semantic fusion learning for job recommendation.

    \item To mitigate the effect of noise in user interactions, we develop a novel graph wavelet kernel-based hypergraph convolutional neural network, which utilizes spectral graph theory to filter noisy data adaptively.

    \item Extensive experiments on three real-world recruitment datasets and a half-week online algorithm deployment on an online recruitment platform demonstrate the effectiveness and efficiency of the framework \name.
\end{itemize}

The remainder of this paper is organized as follows:  In Section \ref{sec:preliminaries}, a crucial definition and the problem statement are discussed. In Section \ref{sec:method}, the proposed framework is introduced. Section \ref{sec:exp} provides the performance evaluation. Section \ref{sec:rw} discusses the related literature, and Section \ref{sec:cc} concludes the paper.

\section{Related Work}
\label{sec:rw}

\subsection{Job Recommender System}

In job recommender systems, various studies have been proposed to match job seekers with recruiters.
As highlighted by \cite{DBLP:journals/kbs/ReusensLBS18,qin2023comprehensive,zheng2024bilateral,DBLP:conf/kdd/HuCZWCZ23,DBLP:conf/aaai/WuQZZC24,zhao2016exploring,DBLP:journals/corr/abs-2302-03525,li2022gromov,li2023automlp,liu2023diffusion,zhao2018recommendations}, while traditional recommender systems are adept at predicting job seekers' preferences, a key to augmenting the system's overall effectiveness lies in addressing the issue of preference drift.
Conventionally, this challenge has been approached through feature engineering in a certain degree, utilizing content data to capture evolving preferences \cite{DBLP:conf/cikm/HuZY18,DBLP:journals/tist/LiHS18,zheng2022cbr}.
Efforts have also been made \cite{DBLP:conf/kdd/HuangWWYSBCJZL21,DBLP:journals/comsis/KannoutGG23,DBLP:conf/kdd/LiuBXG022} to integrate clustering into recommender systems, tackling this problem at the model level by grouping similar users or jobs based on minimal user/job contents.
Although the user and job representations could be enhanced by these refined features, they heavily rely on the results of semantic analysis of fixed content.
% Therefore, we propose a behavioral-semantic fusion framework combining both content and interaction-based approaches to address this problem.
To overcome these limitations, we introduce a behavioral-semantic fusion framework that merges content-driven and interaction-based methodologies, offering a more comprehensive and adaptive solution to the challenge of preference drift. 

\subsection{Recommendation with Graphs}

Graph neural networks are increasingly utilized to capture the complexity of entities and their intricate interrelations. Recent research has leveraged the graph-structured information propagation paradigm to enhance user and item embeddings, employing a variety of neighborhood aggregation techniques \cite{wang2022towards,DBLP:conf/ijcai/Huang21,jiang2023adaptive,DBLP:conf/www/00600ZH023}.
For instance, LightGCN \cite{he2020lightgcn} stands out as a leading graph learning-based recommendation model, celebrated for its streamlined architecture. Drawing inspiration from SGC \cite{wu2019simplifying}, it simplifies the traditional graph convolutional neural networks by omitting transformation matrices and non-linear activation functions, focusing instead on the essential elements of graph convolution.
There are also some studies combining hypergraph learning with recommender systems, including \cite{bu2010music,li2013news,wang2020next,xia2021self}, which use hypergraph to model the short-term user preference for next-item recommendations.
Nonetheless, challenges such as sparsity and noise within the graph can significantly hinder effective information transfer among nodes, potentially confusing the model by prioritizing irrelevant job preferences \cite{dai2021nrgnn,yang2022knowledge}. To counteract these issues, our framework incorporates a wavelet denoising filter, specifically designed to cleanse the data and ensure the accurate extraction of meaningful job preferences, thereby enhancing the overall recommendation process.

\section{Preliminaries}
\label{sec:preliminaries}

In this paper, we adopt the \name\ framework to solve user preference drift during the job-seeking process in top-$k$ job recommendations for users.
Specifically, as shown in the above statistics, we believe a user $u \in \mathcal{U}$ would continue to refine her/his resume along with her/his job preference drift.
% , which would be implied by the job $v \in \mathcal{V}$ that she/he applies for.
Thus, we first segment the user interaction sequence based on the timestamps of resume refinement under the assumption that the job preferences of users remain relatively stable within a given session.

\begin{defi}
\label{def:up}
\textbf{Job Preference Drift}.
It refers to the phenomenon in which users change their job preferences, which could be predominantly observed through whether the user modifies the resume rather than modeling the short-term job-seeking behaviors where user interests tend to remain stable.
\end{defi}

\begin{defi}
\label{def:no}
\textbf{Interaction Noise}.
It stands for job-seeking behaviors that are inconsistent with user preferences, i.e., accidental clicks while a user browses jobs.
\end{defi}

\begin{defi}
\label{def:1}
\textbf{Session-based User-Job Interactions}.
For every user $u$, we need to segment the sequence of the job interactions into multiple sessions $\mathcal{S}^u = \left\{ s^u_1, s^u_2, \cdots \right\}$ based on whether the resume of the user has been changed, in order to obtain more accurate job preference in a specific period.
\end{defi}

Based on the definition above, we formulate the problem of the session-based job recommendation as follows:
\begin{pro}
\label{pro:jobrec}
Given the historical user resumes $\boldsymbol{D}_{t}^u$, job requirements $\boldsymbol{D}_{t}^v$ and interaction sessions $\mathcal{S}^u$,
the objective of the job recommender system is to identify and rank a list of the top-$k$ job vacancies that would likely appeal to a user $u$ during a session $s^{u}$. 
% The system's success hinges on its ability to offer a tailored recommendation for the user by leveraging the accumulated data to discern and extract the user's job preferences.
\end{pro}

\section{Methodology}
\label{sec:method}

\begin{figure*}
    \centering
    \includegraphics[width=0.85\linewidth]{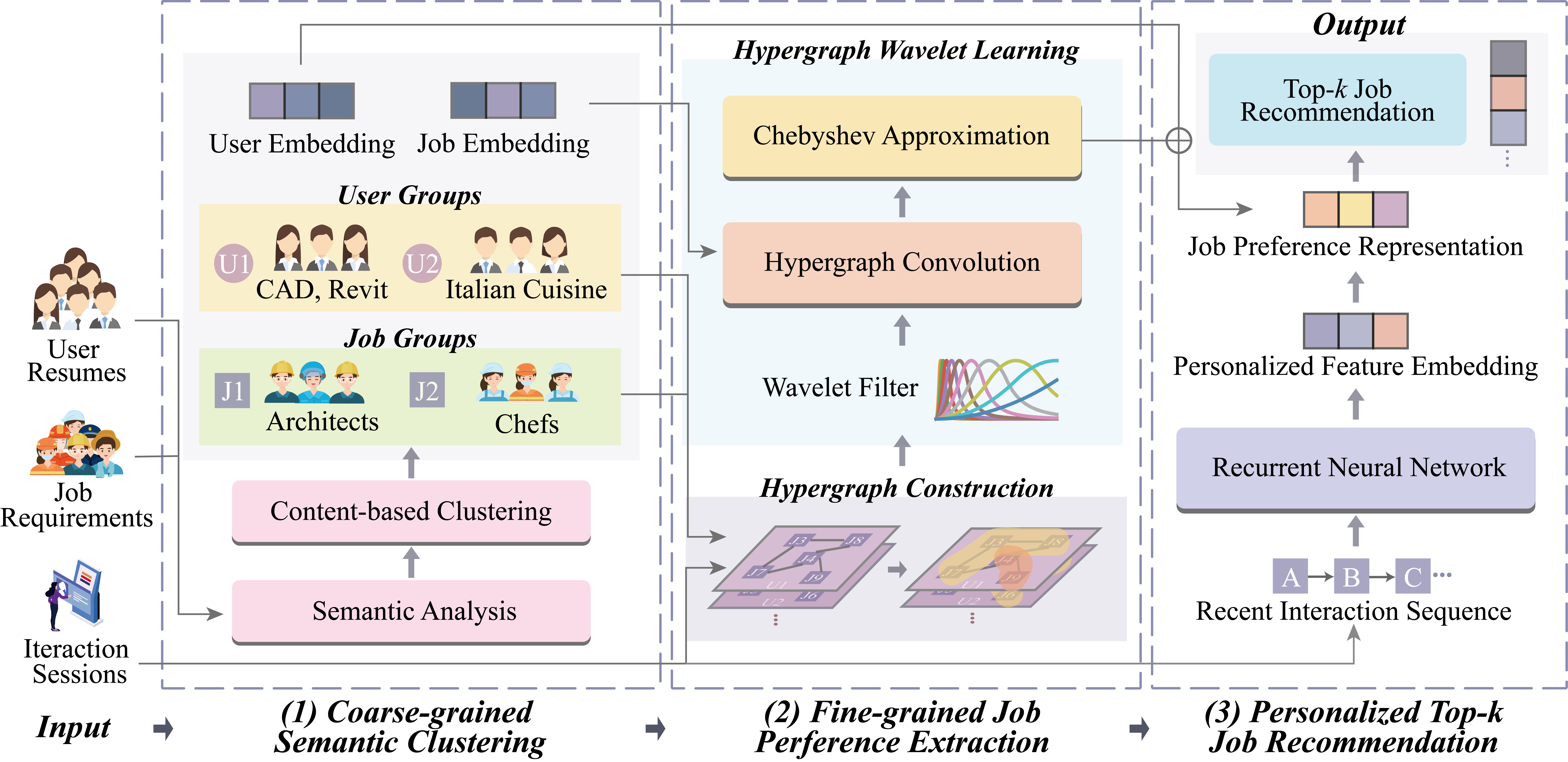}
    \caption{
    The framework overview of \name.
    The framework is divided into three parts: coarse-grained semantic clustering, fine-grained job preference extraction, and personalized top-$k$ job recommendation.
    }
    \label{fig:framework}
\end{figure*}

% In this section, we introduce the proposed framework, \name, as shown in
% Figure \ref{fig:framework}.
% It consists of a coarse-grained feature clustering module, a fine-grained resume content analysis and job preference extraction module, and a personalized recommendation module.
% The first module clusters similar users or jobs into groups based on coarse-grained common features.
% We then conduct a fine-grained analysis of job preferences for user groups with similar refined resume contents.
% In this module, an interaction hypergraph is constructed and a wavelet filtering method on this hypergraph is proposed and applied to enhance data quality and reduce the negative impact of data noise.
% % Finally, a recurrent neural network was applied to extract personalized job preference features based on short-term sequence behavior and combine the above analysis results to recommend the top-$k$ jobs.

In this section, we detail the architecture of the proposed framework, \name, illustrated in Figure \ref{fig:framework}. The framework comprises three primary modules: 1) a coarse-grained semantic clustering module, 2) a fine-grained job preference extraction module, and 3) a personalized top-$k$ recommendation module.

Initially, the coarse-grained semantic clustering module incorporates a feature clustering approach with a probabilistic latent semantic analysis method, which facilitates the identification of broad user or job categories.
The probabilistic latent semantic analysis method could efficiently summarize topics of the resume content and job requirements, and those topics could guide clustering directions.
Subsequent to this, the fine-grained job preference extraction module constructs a multi-granular interaction hypergraph to deal with the data drift issue and then designs an adaptive wavelet learning algorithm for noise-robust preference extraction.
In the hypergraph, we define two types of hyperedges, reflecting the intra-session and inter-session relationships, to introduce more information to the graph.
Moreover, the wavelet filter in hypergraph wavelet learning is designed to detect noise in the spectral domain and further adaptively mitigate the effects of data noise.
The final stage combines a recurrent neural network to discern personalized job preferences from short-term sequential interactions to generate top-$k$ job recommendation results.

%%%%%%%%%%%%%%%%%%%%%%%%%%%%%%%%%%%%%%%%%%%%%%%%%%%%
\subsection{Coarse-grained Semantic Clustering}
\label{sebsec:GFC}
% \zxy{what are the challenges/motivations for this section? Otherwise your technical contribution is trivival}

% Coarse-grained semantic clustering lays the cornerstone for our framework to perform fine-grained preference feature extraction.
% By semantic understanding of job and user, this module enables broad identification of person-job matching.
% As the foundational component of the \name\ framework, we employ a semantic clustering approach that leverages resume and job requirement content to group users and jobs, categorizing them into user groups or occupational types from a coarse-grained perspective.
% This methodology initiates with an in-depth semantic analysis aimed at revealing core themes indicative of job-seeking intentions or recruitment needs.

The coarse-grained semantic clustering serves as the foundational component of our framework, setting the stage for nuanced preference feature extraction. 
By tackling the challenge of aligning diverse and dynamic job preferences with suitable opportunities, this module highlights the core motivation behind the intricate process of facilitating effective employment matching.
% It facilitates a broad yet critical identification of potential matches between job seekers and vacancies through a deep semantic analysis of both jobs and users.
It utilizes semantic insights from resumes and job descriptions to broadly match job seekers with appropriate vacancies, identifying potential fits based on semantic themes related to job preference and recruitment requirements.

% \re{
% The content-based clustering approach for users capitalizes on the distinctions among similar users.
% However, resumes with partially overlapping content do not always indicate congruent job-seeking intentions, presenting a fundamental challenge for content-based clustering analysis.
% Therefore, identifying the primary job-seeking intentions embedded in resume content is crucial for accurately deducing a user's job-seeking trajectory.
% Analogously, clustering jobs based on the thematic similarity of job requirements enables the precise identification of recruitment objectives.
% It will not only refine the matching process between job seekers and vacancies but also streamline the understanding of recruitment needs.
% }

In our model, the conditional probability between the document content $d \in \boldsymbol{D}$ and words $w \in d$ is captured through a latent embedding $z$ ($z= \mathrm{Linear}(w)$ or $\mathrm{Linear}(d)$), representing a class or topic.
The model parameters, $\boldsymbol{P}(w | z)$ and $\boldsymbol{P}(z | d)$, allow for the possibility that words may associate with multiple classes and documents that may cover various topics.
We assume that the distribution of words given a class,
$\boldsymbol{P}(w|z)$ is conditionally independent of the document, implying $\boldsymbol{P}(w|z,d) = \boldsymbol{P}(w|z)$.
Thus, the joint probability of a document $d$ and a word $w$ is represented as:

\begin{equation}
\boldsymbol{P}(w,d) = \boldsymbol{P}(d) \sum_z{\boldsymbol{P}(w|z)\boldsymbol{P}(z|d)}.
\end{equation}

To estimate the parameters $\boldsymbol{P}(w | z)$ and $\boldsymbol{P}(z | d)$, the Expectation-Maximization (EM) algorithm iteratively maximizes the log-likelihood function $\boldsymbol{L}$ over a training corpus $\boldsymbol{D}$:

\begin{equation}
    \boldsymbol{L} = \sum_{d \in \boldsymbol{D}}{\sum_{w \in d}{f(d,w) \log{\boldsymbol{P}(d,w)}}},
\end{equation}
where $f(d,w)$ is the frequency of word $w$ in document $d$.
The EM process alternates between 1) the E-step, estimating the probability $\boldsymbol{P}(z|w,d)$ as:

\begin{equation}
    \boldsymbol{P}(z|w,d) = \frac{\boldsymbol{P}(w|z) \boldsymbol{P}(z|d)}{\sum_{z'}{\boldsymbol{P}(w|z') \boldsymbol{P}(z'|d)}},
\end{equation}
and 2) the M-step, recalculating $\boldsymbol{P}(w | z)$ and $\boldsymbol{P}(z | d)$ to maximize $\boldsymbol{L}$:

\begin{equation}
    \boldsymbol{P}(w|z) = \frac{\sum_d{f(d,w) \boldsymbol{P}(z|w,d)}}{\sum_{w'}\sum_d{f(d,w') \boldsymbol{P}(z|w',d)}},
\end{equation}
\begin{equation}
    \boldsymbol{P}(z|d) = \frac{\sum_w{f(d,w) \boldsymbol{P}(z|w,d)}}{\sum_{z'}\sum_w{f(d,w) \boldsymbol{P}(z'|w,d)}}.
\end{equation}

Following training, the "folding-in" process applies the estimated $\boldsymbol{P}(w|z)$ to test documents $d' \in \boldsymbol{D}$, recalculating $\boldsymbol{P}(z|d')$ while keeping $\boldsymbol{P}(w|z)$ constant.
Typically, only a few iterations of the EM algorithm are required for this process.

After semantic analysis, we combine the normalized document and word latent embeddings $\{z_1, z_2, \cdots\}$ with other normalized attributes ${attr}$ such as age to achieve clustering by K-Means algorithm for the whole data $\boldsymbol{C} := \boldsymbol{C}_\mathcal{U} \text{ or } \boldsymbol{C}_\mathcal{V}$ due to the fact of its high efficiency, as shown in Equation \eqref{eq:kmeans}.

\begin{equation}
\label{eq:kmeans}
    \arg\min_{\boldsymbol{C}} \sum_i^K \sum_{x \in C_i} || x - \mu_i ||_2^2,
\end{equation}
where $\mu_i$ is the mean of all data in $C_i$, $x = \mathrm{concat}(z_1, z_2, \cdots$, $attr)$,
$K$ is the number of clusters.

%%%%%%%%%%%%%%%%%%%%%%%%%%%%%%%%%%%%%%%%%%%%%%%%%%%%
\subsection{Fine-grained Job Preference Extraction}
% \zxy{what are the challenges/motivations for this section? Otherwise your technical contribution is trivival}

% In this module, for effectively capturing the current job preferences of users, we propose a novel spectral graph neural network, which integrates semantic comprehension derived from the first module with session-level short-term interactions.
% Besides, to handle the preference drift and the corresponding noise in short-term interactions, we propose a novel adaptive hypergraph wavelet learning method to address both problems together.

Short-term interactions at a session level always encompass issues of user preference drift and noisy interactions, with each influencing the other. 
To tackle the dual challenges above, the fine-grained job preference extraction module utilizes a novel adaptive hypergraph wavelet learning method in a unified approach.
% This approach highlights our focus on resolving significant difficulties in preference extraction.

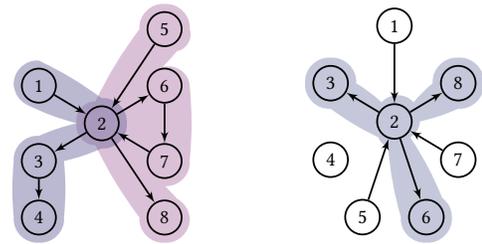
\begin{figure}[htb]
\centering

\tikzstyle{input}=[rectangle, rounded corners, minimum width=1em, minimum height=1em, text centered, draw=black, fill=red!10, drop shadow]
\tikzstyle{other}=[rectangle, rounded corners, minimum width=1em, minimum height=1em, text centered, draw=black, fill=yellow!10, drop shadow]
\tikzstyle{layer}=[rectangle, rounded corners, minimum width=1em, minimum height=1em, text centered, draw=black, fill=blue!10, drop shadow]
\tikzstyle{blank} = [align=center]
\tikzstyle{cnode} = [circle, draw]

\subfigure[Session hyperedges: The pink and grey areas are two hyperedges.]{
\makebox[0.43\linewidth][c]{
\resizebox{0.3\linewidth}{!}{
\begin{tikzpicture}[thick, auto, node distance=1.3cm,>=latex']

\path

(0,0) node [cnode] (node4) {4}
(0,0.9) node [cnode] (node3) {3}
(1,1.5) node [cnode] (node2) {2}
(0,2.1) node [cnode] (node1) {1}

(2,0) node [cnode] (node8) {8}
(2,0.9) node [cnode] (node7) {7}
(2,2.1) node [cnode] (node6) {6}
(2,3) node [cnode] (node5) {5}
;

\draw[->,black,thick] (node3) -- (node4);
\draw[->,black,thick] (node2) -- (node3);
\draw[->,black,thick] (node1) -- (node2);
\draw[->,black,thick] (node5) -- (node2);
\draw[->,black,thick] (node2) -- (node6);
\draw[->,black,thick] (node6) -- (node7);
\draw[->,black,thick] (node7) -- (node2);
\draw[->,black,thick] (node2) -- (node8);

\begin{scope}[on background layer]
    \fill[mycolor42,opacity=0.5] (2, 3.4) 
    to [bend right=50] (1.6,3)
    to [bend right=5] (0.9,1.85) 
    to [bend right=50] (0.6,1.5)
    to [bend right=50] (0.9,1.15)
    to [bend right=5] (1.6,0)
    to [bend right=50] (2,-0.4)
    to [bend right=50] (2.4,0)
    to [bend right=25] (2.2,0.3)
    to [bend right=5] (1.2,1.2)
    to [bend right=5] (1.8,0.7)
    to [bend right=25] (2,0.5)
    to [bend right=50] (2.4,0.9)
    to [bend right=5] (2.4,2.1)
    to [bend right=50] (2,2.4)
    to [bend right=25] (1.8,2.3)
    to [bend right=5] (1.2,1.7)
    to [bend right=5] (2.2,2.65)
    to [bend right=50] (2.4,3)
    to [bend right=50] (2,3.4);

    \fill[mycolor55,opacity=0.5] (0, 2.5) 
    to [bend right=50] (-0.4, 2.1) 
    to [bend right=25] (-0.2,1.8) 
    to [bend right=5] (0.6,1.5)
    to [bend right=5] (0,1.3)
    to [bend right=50] (-0.4, 0.9)
    to [bend right=5] (-0.4,0)
    to [bend right=50] (0, -0.4)
    to [bend right=50] (0.4, 0)
    to [bend right=5] (0.4, 0.9)
    to [bend right=5] (1,1.1)
    to [bend right=50] (1.4,1.5)
    to [bend right=50] (1,1.9)
    to [bend right=5] (0,2.5);
\end{scope}

\end{tikzpicture}
}}
\label{fig:hypa}
}
\subfigure[The transition hyperedge: The grey area is the hyperedge for node 2.]{
\makebox[0.43\linewidth][c]{
\resizebox{0.3\linewidth}{!}{
\begin{tikzpicture}[thick, auto, node distance=1.3cm,>=latex']

\path

(1,3) node [cnode] (node1) {1}
(1,1.5) node [cnode] (node2) {2}
(0,2.1) node [cnode] (node3) {3}
(0,0.9) node [cnode] (node4) {4}
(0.5,0) node [cnode] (node5) {5}
(1.5,0) node [cnode] (node6) {6}
(2,0.9) node [cnode] (node7) {7}
(2,2.1) node [cnode] (node8) {8}
;

\draw[->,black,thick] (node1) -- (node2);
\draw[->,black,thick] (node2) -- (node3);
\draw[->,black,thick] (node5) -- (node2);
\draw[->,black,thick] (node2) -- (node6);
\draw[->,black,thick] (node7) -- (node2);
\draw[->,black,thick] (node2) -- (node8);

\begin{scope}[on background layer]
    \fill[mycolor54,opacity=0.5] (0,2.5) 
    to [bend right=50] (-0.4,2.1)
    to [bend right=50] (0,1.7) 
    to [bend right=5] (0.75,1.25)
    to [bend right=25] (0.9,1.1)
    to [bend right=5] (1.1,0)
    to [bend right=50] (1.5,-0.4)
    to [bend right=50] (1.9,0)
    to [bend right=25] (1.75,0.3)
    to [bend right=5] (1.25,1.25)
    to [bend right=25] (1.4,1.5)
    to [bend right=5] (2,1.7)
    to [bend right=50] (2.4,2.1)
    to [bend right=50] (2,2.5)
    to [bend right=50] (1.6,2.1)
    to [bend right=5] (1.3,1.8)
    to [bend right=50] (0.8,1.8)
    to [bend right=5] (0.4,2.1)
    to [bend right=50] (0,2.5);
\end{scope}

\end{tikzpicture}
}}
\label{fig:hypb}
}
\caption{Two types of hyperedges.}
\label{fig:hyp}
% \vspace{-5mm}
\end{figure}

Initially, employing a standard graph structure to map user-job interactions often results in a proliferation of isolated vertices and edges, adversely impacting the efficacy of graph learning-based job preference extraction. In response, this paper introduces a hypergraph structure, denoted as $\mathcal{G}^{C_u}=(\mathcal{V}^{C_u}, \mathcal{E}^{C_u})$, which utilizes two specialized types of hyperedges to enhance the data with additional insights. The hypergraph for each user group $C_u$ encompasses $n$ job group nodes, alongside corresponding features (graph signals) $\boldsymbol{X}_t^{C_u}$.
The hyperedge, defined as $e = \mathrm{link}( v_a, v_b, \cdots) \in \mathcal{E}^u$, constitutes a subset of the vertex set $\mathcal{V}$, capturing complex, high-order relationships within the graph.
For illustration, consider two sessions: $s_1^{C_u} = \{v_1$, $v_2$, $v_3$, $v_4 \}$, $s_2^{C_u} = \{v_5$, $v_2$, $v_6$, $v_7$, $v_2$, $v_8\}$.
The introduction of two distinct hyperedge types significantly augments the data connectivity within the user-job graph, as depicted in Figure \ref{fig:hyp}.

\noindent \textbf{Session Hyperedges $\mathcal{E}_s^{C_u}$}.
The intra-session relationship is demonstrated as one of the critical factors to session-based recommendation \cite{DBLP:journals/corr/HidasiKBT15}.
For each user group, we link all jobs in each session to enhance the connectivity of these jobs. 
As for the job $v_2$ in Figure \ref{fig:hypa}, we connect the session jobs $\{ v_1, v_3, v_4 \}$ and $\{ v_5, v_6, v_7, v_8 \}$ that include it with a hyperedge, respectively.
It reveals the high-order correlation of jobs facilitating the interaction on $v_2$.

\noindent \textbf{Transition Hyperedges $\mathcal{E}_v$}.
Since the first type of hyperedges cannot model the chronological order of user-job interactions, we utilize job transition hyperedges to address this issue.
For example, we connect the outcoming jobs $\{ v_3, v_6, v_8 \}$ for job $v_2$ as a hyperedge in Figure \ref{fig:hypb}, which also implies the inter-session relationship in those two sessions
% $v_1 \rightarrow v_2 \rightarrow v_3 \rightarrow v_4$ and $v_5 \rightarrow v_2 \rightarrow v_6 \rightarrow v_7 \rightarrow v_2 \rightarrow v_8$ in Figure \ref{fig:hypa}
.
These hyperedges also tackle the issue of preference drift among sessions.

% Moreover, general data-cleaning methods are usually independent of the whole recommendation framework, which leads to the additional overhead of recommender systems.
To sufficiently filter the noise while capturing the job preferences, we design a spectral-based hypergraph wavelet convolutional neural network with a graph filter in the spectral domain for graph convolutional operation.

The graph convolution of the general graph signal $\boldsymbol{x}$ with a filter $\boldsymbol{g} \in \mathbb{R}^n$ is defined as:
\begin{equation}
    \centering
    \begin{aligned}
        \boldsymbol{x} *_{G} \boldsymbol{g} &= \mathcal{F}^{-1}\left(\mathcal{F}(\boldsymbol{x}) \odot \mathcal{F}(\boldsymbol{g}) \right) \\
        &= \boldsymbol{U} \left( \boldsymbol{U}^\mathrm{T} \boldsymbol{x} \odot \boldsymbol{U}^\mathrm{T} \boldsymbol{g}  \right),
    \end{aligned}
\end{equation}
where $*_G$ stands for the convolution operator, $\odot$ denotes the element-wise product. If we denote a filter as $\boldsymbol{g}_\theta = \mathrm{diag}\left( \boldsymbol{U}^\mathrm{T} \boldsymbol{g} \right)$, then the spectral graph convolution can be simplified as:
\begin{equation}
\label{eq_spe_conv}
    \boldsymbol{x} *_{G} \boldsymbol{g}_\theta = \boldsymbol{U} \boldsymbol{g}_\theta \boldsymbol{U}^\mathrm{T} \boldsymbol{x}.
\end{equation}
Almost all spectral-based graph convolutional neural networks follow the definition above, and the key difference lies in the choice of the filter.
For example, Bruna \etal~\cite{bruna2013spectral} take a filter $\boldsymbol{g}_\theta = \Theta^{(k)}_{i,j}$ to be a set of learnable parameters and consider graph signals having multiple channels.
They define the graph convolutional layer as:
\begin{equation}
\label{eq:spcnn}
    \boldsymbol{X}_{:,j}^{(k)} = \sigma \left( \sum_{i=1}^{d_{k-1}}{\boldsymbol{U} \Theta_{i,j}^{(k)} \boldsymbol{U}^\mathrm{T} \boldsymbol{X}_{:,i}^{(k-1)}} \right),
    j = 1,2,\cdots,d_k,
\end{equation}
where $k$ is the layer index,
$\boldsymbol{X}^{(k-1)} \in \mathbb{R}^{n \times d_{k-1}}$ is the input graph signal,
$\boldsymbol{X}^{(0)}_{:,i} = \boldsymbol{x} \in \mathbb{R}^{n \times 1}$,
$d_{k-1}$ is the number of input channels and $d_k$ is the number of output channels,
$\Theta_{i,j}^{(k)}$ is a diagonal matrix filled with learnable parameters.

From Equation \eqref{eq:spcnn}, the Laplacian eigenvectors of the hypergraph need to be precomputed to realize the mapping of hypergraph features between the vertex and spectral domains.
Note that the Laplacian matrix of a hypergraph is defined as follows: $\mathcal{L} := \boldsymbol{D}_v - \boldsymbol{A} = \boldsymbol{D}_v - \boldsymbol{H} \boldsymbol{W} \boldsymbol{D}_e^{-1} \boldsymbol{H}^{\top}$, where $\boldsymbol{A}$ is the adjacency matrix of the hypergraph, $\boldsymbol{H}$ is the node-edge relationship matrix, $\boldsymbol{D}_v$ and $\boldsymbol{D}_e$ are degree matrix of nodes and hyperedges separately. 
Then, the eigenvectors can be obtained by eigendecomposition methods, as shown in Equation \eqref{eq:decomp}.

\begin{equation}
\label{eq:decomp}
\boldsymbol{D}_v - \boldsymbol{H} \boldsymbol{W} \boldsymbol{D}_e^{-1} \boldsymbol{H}^{\top} = \boldsymbol{U} \Lambda \boldsymbol{U}^{\top},
\end{equation}
where $\Lambda$ is a diagonal matrix of Laplacian eigenvalues.

In addition, the filter $\boldsymbol{g}_\theta$ utilized in Equation \eqref{eq:spcnn} is usually a set of learnable hyperparameters but has problems such as convergence difficulty.
Therefore, we utilize the wavelet kernel to finely define a series of filters that filter in-session noise adaptively for different user groups.
Similar to the hypergraph Fourier transform, the hypergraph wavelet transform projects the hypergraph signal from the vertex domain into the spectral domain.
Graph wavelet transform employs a set of wavelets as bases, defined as $\psi_\kappa = \mathrm{concat}(\psi_{1}^\kappa, \psi_{2}^\kappa, \cdots, \psi_{n}^\kappa) \in \mathbb{R}^{n \times n}$, where each wavelet $\psi_{i}^\kappa \in \mathbb{R}^{1 \times n}$ corresponds to a signal on graph diffused away from node $i$ and $\kappa$ is a scaling parameter, which is adapted to spectrum bounds.
Mathematically, $\psi_\kappa$ and $\psi^{-1}_\kappa$ can be written as

\begin{equation}
\label{eq:wavelet_kernel}
\centering
\begin{aligned}
\psi_\kappa = \boldsymbol{U} \boldsymbol{G^\kappa}(\boldsymbol{\Lambda}) \boldsymbol{U}^{\mathrm{T}}, \quad
\psi_\kappa^{-1} = \boldsymbol{U} \boldsymbol{G^\kappa}(\boldsymbol{\Lambda}') \boldsymbol{U}^{\mathrm{T}},
\end{aligned}
\end{equation}
where $G^\kappa(\boldsymbol{\Lambda})=\mathrm{diag} [ g(\kappa\lambda_1), \cdots, g(\kappa_{\lambda_n})]$ and $G^\kappa(\boldsymbol{\Lambda}')=\mathrm{diag} [g^{-1}(\kappa_{\lambda_1})$
,
$\cdots, g^{-1}(\kappa_{\lambda_n})]$ are scaling matrix.

To reduce the computational overhead incurred by the inverse operation, we choose the heat kernel $g(x):= e^{-x}$ as the wavelet mother kernel in this paper, and then $g^{-1}(x) = e^{- (-x)} = g(-x)$.

However, eigen-decomposition is known to have extremely high computational overhead.
In order to avoid the computational cost caused by solving the eigen-decomposition of the Laplacian matrix,
we use Chebyshev polynomials to approximate the convolutional operator $\boldsymbol{x} *_{G} \boldsymbol{g}_\theta \approx \sum_{i=1}^{p}{c_i T_i\left(\boldsymbol{\tilde\Lambda}\right)}$, 
where $\boldsymbol{\tilde{\Lambda}} = \frac{2}{g_{\max}} \boldsymbol{\Lambda}  - \boldsymbol{I}_n$ and $\boldsymbol{\tau} = [\tau_1, \cdots, \tau_{p}]$ is a vector of Chebyshev coefficients.
Note that $T_i(\tilde \psi_\kappa) = \boldsymbol{U} T_i(\boldsymbol{\tilde\Lambda}) \boldsymbol{U}^{\mathrm{T}}$, $\tilde \psi_s = \frac{2}{g_{\max}} \psi_s - \boldsymbol{I}_n$, we have:

\begin{equation}
\label{eq:filter_C}
\psi_{\kappa} \approx P_p(\psi_\kappa) =  \sum_{i=1}^{p} \tau_i T_i(\tilde \psi_\kappa).
\end{equation}

Similarly,

\begin{equation}
\label{eq:filter_inv_C}
\psi_\kappa^{-1} \approx P_p(\psi_\kappa^{-1}) = \sum_{i=0}^{p} \tau_i' T_i(\tilde \psi_\kappa^{-1}),
\end{equation}
where $\tilde \psi_\kappa^{-1} = \frac{2}{{(g^{-1})}_{\max}} \psi_\kappa^{-1} - \boldsymbol{I}_n$ and $\tau_i'$ is a vector of Chebyshev coefficients.

% The Chebyshev polynomials are defined recursively by $T_i(\boldsymbol{x}) = 2 \boldsymbol{x} T_{i-1}(\boldsymbol{x}) - T_{i-2}(\boldsymbol{x})$ with $T_0(\boldsymbol{x}) = \boldsymbol{1}$ and $T_1(\boldsymbol{x}) = \boldsymbol{x}$.

Therefore,
the $l$-th hypergraph wavelet convolutional network for job preference feature extraction can be derived as follows:

\begin{equation}
\label{eq:layer}
\centering
\begin{aligned}
\boldsymbol{X}_{:, j}^{l+1}=
\sigma \left( \sum_{i=1}^{d_{l}} 
  w_{i,j} |\mathcal{V}|^{-1} \sum_{\kappa=1}^{|\mathcal{V}|}{P_p(\psi_\kappa)(\tilde{\mathcal{L}}) \boldsymbol{g}_{\kappa,\theta_{(j,i)}}^l P_p(\psi_\kappa^{-1})(\tilde{\mathcal{L}}) \boldsymbol{X}_{:, i}^l}\right).
\end{aligned}
\end{equation}

Generally, many studies assume the Chebyshev coefficients $\{ \tau_1$, $\tau_2$, $\cdots \}$ in Equation \eqref{eq:filter_C} and \eqref{eq:filter_inv_C} to be a set of random variables that can be trained by a Multi-Layer Perceptron (MLP).
In this paper, we use Boyd's theory \cite{boyd2001chebyshev} to calculate those coefficients directly.
From Boyd's theory, the Chebyshev coefficients can be expressed by:

\begin{equation}
\label{eq:Cc}
\tau_j=\frac{2}{n+1} \sum_{\kappa=0}^n f\left(y_\kappa\right) T_j\left(y_\kappa\right),
\end{equation}
where $y_\kappa := cos\left( \frac{(2\kappa+1)\pi}{2n+2} \right), \kappa \in [n]$ is the interpolant of degree $n$ in the Chebyshev points of the first kind, and $f(\cdot)$ is the Chebyshev series for the function $f$ that absolutely continuous on $[-1, 1]$, which is defined as \cite{sloan1980product, trefethen2008gauss}:

\begin{equation}
\label{eq:Cs}
   f(x)=\sum_{j=0}^{\infty} b_j T_j(x), \quad b_j=\frac{2}{\pi} \int_{-1}^1 \frac{f(x) T_j(x)}{\sqrt{1-x^2}} d x ,
\end{equation}
where the prime denotes a sum whose first term is halved and $T_j(x)=\cos \left(j \cos ^{-1} x\right)$ is the Chebyshev polynomial of degree $j$.

The Chebyshev coefficient $\tau_j$ in Equation \eqref{eq:Cc} can be computed by Fast Fourier Transform (FFT), which significantly improves computational efficiency compared to eigen-decomposition methods.

%%%%%%%%%%%%%%%%%%%%%%%%%%%%%%%%%%%%%%%%%%%%%%%%%%%%
\subsection{Personalized Top-$k$ Job Recommendation}
% \zxy{what are the challenges/motivations for this section? Otherwise your technical contribution is trivival}

% To ensure a precise match between users' job preferences identified in the last section and the recommended jobs, 
% % to meticulously customize the correlation of job preferences with recommended positions for each user
% After getting a comprehensive current job preference representation, \name\ employs a recurrent neural network for job recommendations, drawing on recent user interactions for enhanced accuracy.

Adapting job recommendations to reflect the job preference of the current user is the final key challenge faced by job recommendation systems, which could directly affect the recommendation performance.
To address this difficulty, \name\ employs a recurrent neural network to align recommendations with previously extracted job preferences precisely. This approach underscores our dedication to providing timely and relevant job matches, ensuring high accuracy in meeting user needs.

To refine the analysis of a user's personalized features, we consider the last $T$ jobs with which a user has recently interacted, along with their embeddings $\boldsymbol{Emb}_t^{\text{job}}, t \in [T]$.
A recurrent neural network is employed to generate the personalized feature $\boldsymbol{Y}_t$:

\begin{equation}
\label{eq:rnn}
\begin{aligned}
\boldsymbol{Y}_t &= \sigma(\boldsymbol{W}_a \cdot \boldsymbol{O}_t), \\
\boldsymbol{O}_t &= \sigma(\boldsymbol{W}_b \cdot \boldsymbol{Emb}_t^{\text{job}} + \boldsymbol{W}_c \cdot \boldsymbol{O}_{t-1}),
\end{aligned}
\end{equation}
where $\boldsymbol{O}_{0} = \boldsymbol{0}$, $\sigma$ is the activation function, $\boldsymbol{W}_a$, $\boldsymbol{W}_b$ and $\boldsymbol{W}_c$ are learnable parameters.

Then, this personalized feature enhances the job preference $\boldsymbol{X}^{L}$ calculated by Equation \eqref{eq:layer} for a specific user of a certain feature by the following weighted formula:

\begin{equation}
    \boldsymbol{Y} = \text{Sigmod}\left(\text{Linear}\left( \text{concat}(\boldsymbol{Y}_T, \boldsymbol{X}^{L})\right)\right).
\end{equation}

\name\ gives the top-$k$ job recommendation by minimizing the loss function below:

\begin{equation}
\label{eq:loss}
Loss = \frac{1}{k} \sum_{i=1}^k \sum_{q=1}^{M} \chi_{iq} \log (y_{iq}),
\end{equation}
where $\chi_{iq} \in [0,1]$ is the ground truth, $y_{iq}$ is the value on $i$-th row and $q$-th column in $\boldsymbol{Y}$, standing for the probability that the $i$-th predicted job belongs to job label $q$, $M$ is the number of job positions.

\section{Experiments}
\label{sec:exp}

In this section, we first describe the datasets used in this paper.
Then, we introduce the experimental settings and compare \name\ with representative baselines.
We further present some case studies on job recommendations (\emph{Appendix \ref{app:case}}).
The experiments are mainly designed to answer the research questions as follows:

\noindent $\bullet$ \textbf{RQ1}: Can our \name\ recommend suitable jobs for users?

\noindent $\bullet$ \textbf{RQ2}: Does the clustering module effectively accommodate new jobs or users who have just revised their resumes?

\noindent $\bullet$ \textbf{RQ3}: How does the specially designed hypergraph wavelet learning method deal with preference drift and noise issues?

\noindent $\bullet$ \textbf{RQ4}: How do different settings influence the model performance?

\subsection{Dataset}

The datasets come from the real-world online recruitment markets of multiple cities (Shenzhen, Shanghai, and Beijing).
We utilize the user-job interaction (browse, click, chat, and so on) logs, user resumes, and job requirements data on this platform from July 1, 2023 to January 31, 2024 (215 days in total).
To protect the privacy of users and platform operators, all sensitive information related to users is hashed or removed,
%Since the average daily amount of data generated by the platform exceeds 200,000,000 records, 
and we only keep those data in the information technology industry.
% We will publicize these datasets upon the acceptance to foster the research on this important topic.
% filtered out users who have modified their resumes and also restricted all job positions in the technical category where users intend to work as training samples for experiments.
% For a more comprehensive study on the performance of different models, we split the datasets into several subsets by taxonomy of career locations, \ie Beijing, Shanghai, Guangzhou, and Shenzhen.
The detailed statistical information of our datasets is summarized in Table \ref{tab:datasets}. 

\begin{table}[t]
 \caption{Statistics of the datasets.}
 \label{tab:datasets}
\renewcommand\tabcolsep{5pt}
 \begin{filecontents*}{data/dataset.txt}
City; UserNumber; InteractionRecord; JobNumber
Shenzhen; 9,586; 9,504,285; 1,804,402
Shanghai; 10,233; 8,894,726; 1,707,248
Beijing; 15,915; 13,467,252; 2,012,972
\end{filecontents*}
\csvreader[
 separator=semicolon, % semicolon is ; hash is #
   tabular=cccc,
   table head=\hline
  
   \specialrule{0em}{1pt}{1pt}
   
   % \multicolumn{1}{c}{} & \multicolumn{2}{c}{\bfseries{Train}} & \multicolumn{2}{c}{\bfseries{Valid}} & \multicolumn{2}{c}{\bfseries{Test}} \\
  
   % \specialrule{0em}{1pt}{1pt}

   % \multicolumn{1}{c}{} & \multicolumn{1}{c}{\bfseries{Record}} & \multicolumn{1}{c}{\bfseries{Period}} & \multicolumn{1}{c}{\bfseries{Record}} & \multicolumn{1}{c}{\bfseries{Period}} & \multicolumn{1}{c}{\bfseries{Record}} & \multicolumn{1}{c}{\bfseries{Period}} \\
  
   % \specialrule{0em}{1pt}{1pt}

   \multicolumn{1}{c}{} & \multicolumn{1}{c}{\bfseries{Users}} & \multicolumn{1}{c}{\bfseries{Interactions}} & \multicolumn{1}{c}{\bfseries{Jobs}} \\
  
   \specialrule{0em}{1pt}{1pt}
  
   \hline,
   late after line={\\},
   late after last line= \\
   \hline % horizontal line at the end of the table
 ]{data/dataset.txt}{}{\csvlinetotablerow}
 \end{table}

\subsection{Experimental Settings}

\noindent \textbf{Baselines} \quad
\
We compare \name\ with baselines from different types of recommendation methods, including conventional methods: BasicMF \cite{koren2009matrix}, ItemKNN \cite{wang2006unifying}, PureSVD \cite{cremonesi2010performance}, and SLIM \cite{ning2011slim}, DAE \cite{wu2016collaborative}, MultVAE \cite{liang2018variational}, EASE \cite{steck2019embarrassingly};
Graph neural networks-based methods: SLRec \cite{yao2021self} and SGL \cite{wu2021self}, P3a \cite{cooper2014random}, RP3b \cite{paudel2016updatable}, NGCF \cite{wang2019neural}, LightGCN \cite{he2020lightgcn}, GCCF \cite{chen2020revisiting}, NCL \cite{lin2022improving}, DirectAU \cite{wang2022towards}, HG-GNN \cite{pang2022heterogeneous}, A-PGNN \cite{zhang2020personalized}, AdaGCL \cite{jiang2023adaptive}, and MvDGAE \cite{zheng2021multi}; Sequential recommendation methods:  STAMP \cite{DBLP:conf/kdd/LiuZMZ18}, GRU4Rec \cite{DBLP:journals/corr/HidasiKBT15}, BERT4Rec \cite{DBLP:conf/cikm/SunLWPLOJ19}, CL4Rec \cite{xie2022contrastive}, CoScRec \cite{DBLP:journals/corr/abs-2108-06479}, and TiCoSeRec \cite{DBLP:conf/aaai/DangYGJ0XSL23}.
More details about these baselines are shown in \emph{Appendix \ref{app:baselines}}.
For ablation studies, we compare the variants of \name\ to verify the effectiveness of each component.

\noindent \textbf{Evaluation metrics} \quad
\
In this paper, two metrics commonly used in recommendation algorithms are used as evaluation metrics: hit ratio and mean reciprocal rank, and the definitions of these metrics are demonstrated as follows:

\begin{itemize}[leftmargin=*]
    \item \textbf{Hit Ratio (HR)}: It measures the proportion of successful recommended jobs out of all the recommendations made. Later, we use \textbf{H@$k$} to denote the value of HR when the model makes top-$k$ recommendations.

    \item \textbf{Mean Reciprocal Rank (MRR)}: It is a statistical measure that focuses explicitly on the rank of the first relevant item in the list of recommendations to show the effectiveness of a recommendation method, and in this paper we use the symbol \textbf{M@$k$} to present this metric for simplicity.

\end{itemize}

\noindent \textbf{Implementation Details} \quad 
\
We experiment with a Spark cluster for preprocessing the data and A800 GPU servers to train and infer the proposed model.
It has three parts: coarse-grained semantic clustering, fine-grained job preference extraction, and personalized top-$k$ job recommendation.
1) Coarse-grained semantic clustering: we set the ratio of the number of groups to the number of users to about 1:1000 and the ratio of the number of groups to the number of job numbers to about 1:500.
User professional skills and their inherent characteristics, such as work experience, are used as user clustering characteristics.
Similarly, we extract job requirements as clustering features for jobs.
2) Fine-grained job preference extraction: the sparse matrix is used to represent the structure of hypergraphs efficiently.
In addition, the order of Chebyshev approximation is $p=3$, the total degree of interpolants is $n=50$, and the number of hypergraph convolutional layers is $L = 1$. We set the number of hidden dimensions of this network to $d_l = 128$.
4) Personalized top-$k$ job recommendation: the number of hidden dimensions of the recurrent neural network is set to be the same as the one in hypergraph convolutional layers: $d_{\text{rnn}} = 128$.
We set $k=10$ for baseline experiments, and more experimental results under different settings of $k$ can be found in \emph{Appendix \ref{app:para_k}}.
As for each dataset, we split the training and validation data at a ratio of 4:1, and we randomly sample 20\% data of the whole dataset for testing. 
In addition, Adam is used as the optimizer for all models, and we use the default parameter for it, \ie $\beta_{1}=0.9$, $\beta_2=0.999$, $\epsilon=10^{-8}$, $lr=10^{-3}$.

\subsection{Overall Performance \emph{(RQ1)}}

The performance of all the baselines in three datasets is shown in Table \ref{tab:baselines}, in terms of the two metrics, \ie H@10 and M@10.
The performance of all methods is the average of the last 100 epochs in a total of 1000 epochs. It can be observed:

\begin{table}[!ht]
\caption{Experimental results of different baselines.}
\label{tab:baselines}
\renewcommand\tabcolsep{2.5pt}
\begin{filecontents*}{data/baseline.txt}
title,col1,col2,col3,col4,col5,col6
BasicMF,0.2791,0.2837,0.3528,0.2773,0.3181,0.3004
ItemKNN,0.2479,0.2969,0.3278,0.2793,0.3224,0.2931
PureSVD,0.4422,0.4205,0.4658,0.4391,0.4666,0.3827
SLIM,0.045,0.0443,0.0463,0.047,0.0539,0.04
DAE,0.1867,0.2024,0.1988,0.2125,0.2013,0.2024
MultVAE,0.4249,0.4149,0.4516,0.3499,0.5229,0.348
EASE,0.2891,0.3192,0.2683,0.2989,0.29,0.2818
SLRec,0.2767,0.3796,0.3636,0.2958,0.3968,0.3526
SGL,0.4089,0.3618,0.4375,0.397,0.4128,0.3618
P3a,0.3185,0.3748,0.3177,0.3355,0.3728,0.3489
RP3b,0.2917,0.3017,0.3362,0.2929,0.3304,0.2919
NGCF,0.3557,0.2859,0.398,0.2924,0.3851,0.2807
LightGCN,0.3785,0.3647,0.4423,0.3453,0.4232,0.3911
GCCF,0.3756,0.346,0.3901,0.2993,0.3896,0.3283
NCL,0.3424,0.3917,0.3476,0.3472,0.4111,0.3557
DirectAU,0.3725,0.37,0.4173,0.3759,0.3943,0.3643
HG-GNN,0.3131,0.1936,0.3637,0.2826,0.4204,0.3430
A-PGNN,0.3372,0.2069,0.3876,0.3019,0.4485,0.3657
AdaGCL,0.3867,0.4417,0.4044,0.3814,0.4516,0.3867
MvDGAE,\underline{0.5028},\underline{0.5258},\underline{0.5316},\underline{0.4812},0.4677,0.4845
STAMP,0.2754,0.2924,0.3031,0.2756,0.3073,0.2745
GRU4Rec,0.3247,0.3449,0.3576,0.3251,0.3623,0.3237
BERT4Rec,0.3395,0.3607,0.3738,0.3399,0.3787,0.3385
CL4Rec,0.3882,0.4124,0.4276,0.3887,0.4332,0.3871
CoScRec,0.4096,0.4352,0.4512,0.41,0.457,0.4084
TiCoSeRec,0.4397,0.4671,0.4842,0.4402,\underline{0.5324},\underline{0.4925}
\textbf{Ours},\textbf{0.5598},\textbf{0.5467},\textbf{0.6166},\textbf{0.5152},\textbf{0.6247},\textbf{0.5131}
\textbf{Improves},10.19\%,3.82\%,13.78\%,6.61\%,14.7\%,4.01\%
\end{filecontents*}
\csvreader[
  tabular=ccccccc,
  table head=\hline
  
  \specialrule{0em}{1pt}{1pt}

  \multicolumn{1}{c}{} & \multicolumn{2}{c}{\bfseries{Shenzhen}} & \multicolumn{2}{c}{\bfseries{Shanghai}} & \multicolumn{2}{c}{\bfseries{Beijing}}
  
  \\ \specialrule{0em}{1pt}{1pt}

  \multicolumn{1}{c}{} & \multicolumn{1}{c}{\bfseries{H@10}} & \multicolumn{1}{c}{\bfseries{M@10}} & \multicolumn{1}{c}{\bfseries{H@10}} & \multicolumn{1}{c}{\bfseries{M@10}} & \multicolumn{1}{c}{\bfseries{H@10}} & \multicolumn{1}{c}{\bfseries{M@10}}
  
  \\ \specialrule{0em}{1pt}{1pt}
  
  \hline,
  late after line={\\},
  late after last line= \\
  \hline % horizontal line at the end of the table
]{data/baseline.txt}{}{

\ifnumequal{\thecsvrow}{8}{ \hline }{}
\ifnumequal{\thecsvrow}{21}{ \hline }{}
\ifnumequal{\thecsvrow}{27}{ \hline }{}
\ifnumequal{\thecsvrow}{28}{ \hline }{}
\csvlinetotablerow
}

\small{
Bold indicates the statistically significant improvements \\ (\ie two-sided t-test with p < 0.05) over the best baseline (underlined).

For all metrics: the higher, the better.
}

\end{table}

\begin{itemize}[leftmargin=*]
    \item We can see that most conventional methods have poor performance, \ie BasicMF, ItemKNN, SLIM, DAE, and EASE.
SLIM deploys only a linear function to model user-job interactions, which limits the ability to generalize the model. BasicMF, ItemKNN, DAE, and EASE cannot provide fine-grained modeling for user/job features. PureSVD and MultVAE offer significant enhancements in performance over other techniques, yet they require extensive computational resources. Despite their advantages, they fall short of accurately capturing the dynamics of drifted interactions.
    \item Compared to GNN-based methods, \name\ allows for the extraction of fine-grained user representations from user resumes and their use in preference analysis.
Among these baselines, MvDGAE achieves the best performance.
This is because it also uses noise reduction representation learning based on multiview graphs. However, it lacks content modeling of user resumes and job requirements, resulting in lower results than our framework.
    \item Sequential models can effectively learn the relationship among user-job interactions over time, but such interactions can easily be negatively impacted by spontaneous user preference drift, which would be directly reflected in interaction records.
% in response to different job search needs in real time.
Thus, the sequence-only models do not apply to the job recommendation scenario, and their performance is justifiably worse than that of our framework.
\end{itemize}

Furthermore, we extended our evaluation by deploying our model from the offline experiments to an online recruitment platform for a half-week online experiment, as shown in 
Table \ref{tab:exp_online}.
In online experiments, the performance is measured on a daily basis, and we randomly select a unique 1\% of active users and push the results for each model directly at the re-rank stage.
It demonstrates our proposed method's superior performance and exceptional robustness compared to other baseline models.

\begin{table}[t]
\caption{Results of online experiments.}
\label{tab:exp_online}
\renewcommand\tabcolsep{3pt}
\begin{filecontents*}{data/exp_online.txt}
model,c1,s1,c2,s2,c3,s3,c4,s4
MvDGAE,0.67,0.46,0.79,0.49,0.76,0.53,0.69,0.47
TiCoSeRec,0.45,0.23,0.66,0.41,0.73,0.72,0.70,0.78
CoScRec,0.42,0.30,0.43,0.28,0.46,0.32,0.55,0.46
\textbf{\name},\textbf{0.82},\textbf{0.71},\textbf{1.00},\textbf{0.96},\textbf{0.98},\textbf{1.00},\textbf{0.97},\textbf{0.95}
\end{filecontents*}
\csvreader[
  tabular=ccccccccc,
  table head=\hline
  
  \specialrule{0em}{1pt}{1pt}
  
  \multicolumn{1}{c}{} & \multicolumn{2}{c}{\bfseries{Day 1}} & \multicolumn{2}{c}{\bfseries{Day 2}} & \multicolumn{2}{c}{\bfseries{Day 3}} & \multicolumn{2}{c}{\bfseries{Day 4}}
  \\
  
  \specialrule{0em}{1pt}{1pt}
  
  \multicolumn{1}{c}{} & \multicolumn{1}{c}{\bfseries{C}} & \multicolumn{1}{c}{\bfseries{S}} & \multicolumn{1}{c}{\bfseries{C}} & \multicolumn{1}{c}{\bfseries{S}} & \multicolumn{1}{c}{\bfseries{C}} & \multicolumn{1}{c}{\bfseries{S}} & \multicolumn{1}{c}{\bfseries{C}} & \multicolumn{1}{c}{\bfseries{S}}
  
  \\ \specialrule{0em}{1pt}{1pt}
  \hline,
  late after line={\\},
  late after last line= \\
  \hline % horizontal line at the end of the table
]{data/exp_online.txt}{}{\csvlinetotablerow}

\small{
“\textbf{C}” indicates the rate of having a chat about the recommended jobs, \\
“\textbf{S}” stands for the rate of onboarding to the recommended jobs. \\
* Please note that all results have been \textbf{\textit{normalized}} to \\ safeguard the company's trade secrets.

For all metrics: the higher, the better.
}

\end{table}

\subsection{Ablation Study \emph{(RQ2, RQ3)}}

\noindent \textbf{The influence of clustering} \quad
\
As mentioned previously, the clustering module can effectively achieve semantic matching at a coarse-grained level.
Therefore, we partition the dataset into four subsets that do not overlap each other to train the model to achieve the following four tasks:
1) existing jobs for existing users, 2) existing jobs for users who have just revised resumes, 3) new jobs for existing users, and 4) new jobs for users who have just revised resumes.
The comparison results are shown in Table \ref{tab:clustering}.

\begin{table}[!ht]
\caption{Results under different tasks of the dataset.}
\label{tab:clustering}
\renewcommand\tabcolsep{3pt}
\begin{filecontents*}{data/exp_clustering.txt}
x,col1,col2,col3,col4
Task 1,0.7241,0.5947,0.7119,0.6012
Task 2,0.6889,0.5659,0.2215,0.1819
Task 3,0.6557,0.5385,0.2018,0.1658
Task 4,0.6247,0.5131,0.0855,0.0702
\end{filecontents*}
\csvreader[
  tabular=ccccc,
  table head=\hline
  
  \specialrule{0em}{1pt}{1pt}
  
  \multicolumn{1}{c}{} & \multicolumn{2}{c}{\bfseries{Ours}} & \multicolumn{2}{c}{\bfseries{Non-clustering\ }} 
  % & \multicolumn{2}{c}{\bfseries{Prediction Accuracy (\%)}}
  \\
  
  \specialrule{0em}{1pt}{1pt}
  
  \multicolumn{1}{c}{} & \multicolumn{1}{c}{\bfseries{H@10}} & \multicolumn{1}{c}{\bfseries{M@10}} & \multicolumn{1}{c}{\bfseries{H@10}} & \multicolumn{1}{c}{\bfseries{M@10}} 
  
  \\ \specialrule{0em}{1pt}{1pt}
  \hline,
  late after line={\\},
  late after last line= \\
  \hline % horizontal line at the end of the table
]{data/exp_clustering.txt}{}{\csvlinetotablerow}
\end{table}

Compared to the framework without a clustering module, \name\ is better adapted to semantic matching: neither the HR nor the MRR suffers from a significant drop, which shows the efficiency of the clustering module.

\noindent \textbf{The effect of hyperedges} \quad
\
The construction of hyperedges in \name\ can increase the density of the interaction graph, which could add more structural information to address the preference drift issue.
To verify this idea, we use the following formula to compute the density of an undirected graph: $\textbf{Density} = \frac{2|\mathcal{E}|}{|\mathcal{V}| \times \left(|\mathcal{V}| - 1\right)}$.
As shown in Table \ref{tab:hyperedge}, the density of the graph is doubled when we add all types of hyperedges and the experimental results have also been improved due to the optimization of the structure in the graph.

\begin{table}[!ht]
\caption{Results under different graph constructions.}
\label{tab:hyperedge}
\renewcommand\tabcolsep{3pt}
\begin{filecontents*}{data/exp_hyperedge.txt}
name, sparity, HR, MRR
Without hyperedges,1.177\%,0.4109,0.3375
Only session hyperedges,2.811\%,0.4904,0.4028
Only transition hyperedges,2.793\%,0.4927,0.4046
Both types of hyperedges,3.581\%,0.6247,0.5131
\end{filecontents*}
\csvreader[
  tabular=cccc,
  table head=\hline
  
  % \specialrule{0em}{1pt}{1pt}
  
  % \multicolumn{1}{c}{} & \multicolumn{2}{c}{\bfseries{Ours}} & \multicolumn{2}{c}{\bfseries{Non-clustering}} 
  % % & \multicolumn{2}{c}{\bfseries{Prediction Accuracy (\%)}}
  % \\
  
  \specialrule{0em}{1pt}{1pt}

  \multicolumn{1}{c}{} & \multicolumn{1}{c}{\bfseries{Density}} & \multicolumn{1}{c}{\bfseries{H@10}} & \multicolumn{1}{c}{\bfseries{M@10}}
  
  \\ \specialrule{0em}{1pt}{1pt}
  \hline,
  late after line={\\},
  late after last line= \\
  \hline % horizontal line at the end of the table
]{data/exp_hyperedge.txt}{}{\csvlinetotablerow}
\end{table}

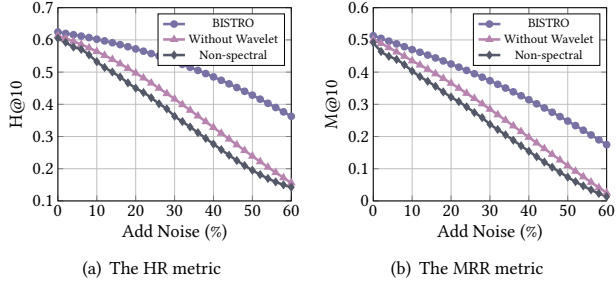
\begin{figure}[]
\centering
\subfigure[The HR metric]{
\label{fig:exp_filter_line_a}
\resizebox{0.476\linewidth}{!}{
\begin{tikzpicture}
\begin{axis}[
	smooth,
	grid=major,
	xlabel=Add Noise (\%),
	ylabel=H@10,
	xmin=0,
	xmax=60,
        ymin=10,
        ymax=70,
	xtick = {0,10,20,30,40,50,60},
	xticklabels={0,10,20,30,40,50,60},
	legend columns=1,
	ytick = {0,10,20,30,40,50,60,70},
	yticklabels={0,0.1,0.2,0.3,0.4,0.5,0.6,0.7},
	legend style={font=\large,at={(0.7,0.7)},anchor=south},
	font=\huge
	]
	\addplot [each nth point=2, mark=*,mycolor55, line width=2pt] table [x=x, y=ours,, col sep=comma] {data/exp_filter_HR.txt};
    \addplot [each nth point=2, mark=triangle*,mycolor42, line width=2pt] table [x=x, y=nonwavelet,, col sep=comma] {data/exp_filter_HR.txt};
    \addplot [each nth point=2, mark=diamond*,mycolor51, line width=2pt] table [x=x, y=nonspectral,, col sep=comma] {data/exp_filter_HR.txt};
	\legend{\name, Without Wavelet, Non-spectral}
\end{axis}
\end{tikzpicture}
}}
\subfigure[The MRR metric]{
\label{fig:exp_filter_line_b}
\resizebox{0.476\linewidth}{!}{
\begin{tikzpicture}
\begin{axis}[
	smooth,
	grid=major,
	xlabel=Add Noise (\%),
	ylabel=M@10,
	xmin=0,
	xmax=60,
        ymin=0,
        ymax=60,
	xtick = {0,10,20,30,40,50,60},
	xticklabels={0,10,20,30,40,50,60},
	ytick = {0,10,20,30,40,50,60},
	yticklabels={0,0.1,0.2,0.3,0.4,0.5,0.6},
	legend columns=1,
	legend style={font=\large,at={(0.7,0.7)},anchor=south},
	font=\huge
	]
	\addplot [each nth point=2, mark=*,mycolor55, line width=2pt] table [x=x, y=ours,, col sep=comma] {data/exp_filter_MRR.txt};
    \addplot [each nth point=2, mark=triangle*,mycolor42, line width=2pt] table [x=x, y=nonwavelet,, col sep=comma] {data/exp_filter_MRR.txt};
    \addplot [each nth point=2, mark=diamond*,mycolor51, line width=2pt] table [x=x, y=nonspectral,, col sep=comma] {data/exp_filter_MRR.txt};
	\legend{\name, Without Wavelet, Non-spectral}
\end{axis}
\end{tikzpicture}
}}
\caption{Results of model performance in relation to the proportion of noise in data.}
\label{fig:exp_filter_line}
\end{figure}

\input{draw/exp_filter_heat}

\begin{figure*}[t]
\centering
\subfigure[Different user group sizes]{
    \label{fig:para_user}
    \makebox[0.27\linewidth][c]{
    \resizebox{0.24\linewidth}{!}{
        \begin{tikzpicture}
            \begin{axis}[
                ybar,
                grid=major,
        	xlabel=$\xi_u$, %Ratio of users to user groups 
                ymin=0.3,
                ymax=0.7,
                bar width=16pt,
                % nodes near coords,
                enlarge x limits=0.3,
                xtick = {1,2,3},
                xticklabels={500:1,1000:1,2000:1},
                % ylabel=Value,
                legend columns=2,
                legend style={font=\large},
                font=\huge
                ]
                \addplot [fill=mycolor55light, draw=mycolor55, postaction={pattern=crosshatch,pattern color=mycolor55}] table [x=x,y=HR,,col sep=comma] {data/exp_para_user.txt};
                \addplot [fill=mycolor42light2, draw=mycolor42, postaction={pattern=grid, pattern color=mycolor42}] table [x=x,y=MRR,,col sep=comma] {data/exp_para_user.txt};
                \legend{H@10, M@10}
            \end{axis}
        \end{tikzpicture}
    }}
}
\subfigure[Different job group sizes]{
    \label{fig:para_job}
    
\makebox[0.27\linewidth][c]{
    \resizebox{0.24\linewidth}{!}{
        \begin{tikzpicture}
            \begin{axis}[
                ybar,
                grid=major,
        	xlabel=$\xi_v$, %Ratio of jobs to job groups 
                ymin=0.3,
                ymax=0.7,
                bar width=16pt,
                % nodes near coords,
                enlarge x limits=0.3,
                xtick = {1,2,3},
                xticklabels={100:1,500:1,1000:1},
                % ylabel=Value,
                legend columns=2,
                legend style={font=\large},
                font=\huge
                ]
                \addplot [fill=mycolor55light, draw=mycolor55, postaction={pattern=crosshatch,pattern color=mycolor55}] table [x=x,y=HR,,col sep=comma] {data/exp_para_job.txt};
                \addplot [fill=mycolor42light2, draw=mycolor42, postaction={pattern=grid, pattern color=mycolor42}] table [x=x,y=MRR,,col sep=comma] {data/exp_para_job.txt};
                \legend{H@10, M@10}
            \end{axis}
        \end{tikzpicture}
}}}
\subfigure[Different orders of Chebyshev approximation]{
\label{fig:para_p}
\makebox[0.27\linewidth][c]{
\resizebox{0.24\linewidth}{!}{
    \begin{tikzpicture}
        \begin{axis}[
            ybar,
            grid=major,
            ymin=0.25,
            ymax=0.7,
            bar width=10pt,
            enlarge x limits=0.1,
            xtick = {1,2,3,4,5},
            xticklabels={1,2,3,4,5},
            % nodes near coords,
            % ylabel=Value,
            xlabel=$p$,
            legend columns=2,
            legend style={font=\large},
            font=\huge
            ]
            \addplot [fill=mycolor55light, draw=mycolor55, postaction={pattern=crosshatch,pattern color=mycolor55}] table [x=p,y=HR,,col sep=comma] {data/exp_para_p.txt};
            \addplot [fill=mycolor42light2, draw=mycolor42, postaction={pattern=grid, pattern color=mycolor42}] table [x=p,y=MRR,,col sep=comma] {data/exp_para_p.txt};
            \legend{H@10, M@10}
        \end{axis}
    \end{tikzpicture}
}}}
\centering
\caption{Results of different hyperparameters.}
\label{fig:para}
\end{figure*}
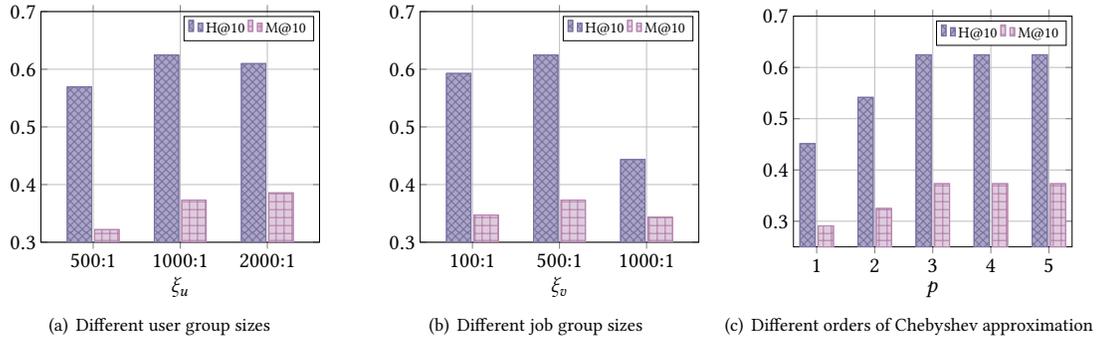

\noindent \textbf{The validity of hypergraph wavelet filter} \quad
\
In \name, we design a novel hypergraph wavelet learning method. 
In this learning method, a wavelet filter is deployed for data denoising as well as fine-grained job preference feature extraction.
As shown in Figure~\ref{fig:exp_filter_line}, the curves illustrate the results of three models, which have different filtering settings, under different percentages of noise in the data.
We can also visualize from it that our method, \name, has the smoothest decrease in model performance as the proportion of noise in the data increases.

%Notably, as data noise levels escalated, the comparative models demonstrated diminished noise filtering effectiveness relative to our proposed approach. Specifically, the random walk-based method significantly underperformed compared to the spectral GCN method, primarily due to spectral graph neural networks' ability to filter out irrelevant interaction features. Furthermore, our approach employs a wavelet kernel to create a set of sub-filters, adeptly denoising by dynamically selecting appropriate filters for the user's evolving characteristics.

To vividly show the denoising capability of the proposed hypergraph wavelet filter, we randomly select a user who is active in a week, filter the 50 most recent interactions from three job categories, and construct an interaction graph.
In this graph, each node represents a job the user has engaged with, interconnected by grey dotted lines, while the interaction sequence of the user is depicted with grey edges.
On this basis, we introduce noisy jobs (marked with orange crosses) and their corresponding interactions (denoted by orange edges and dotted lines) to mimic the effect of a user accidentally clicking on unrelated job types.
Given that each model generates job preference representations for diverse jobs, we visualize the connections between the user and jobs, as well as the relationships among jobs themselves, as shown in Figure~\ref{fig:exp_filter}.
We eliminate edges whose cosine similarity between job representation pairs fell below a uniform threshold and remove links between isolated jobs and the user.
Consequently, a graph with more orange lines indicates lower model performance.
Notably, as data noise levels escalated, the comparative models demonstrated diminished noise filtering effectiveness relative to our proposed approach. Specifically, the random walk-based method significantly underperformed compared to the spectral GCN method, primarily due to the ability of spectral graph neural networks to filter out irrelevant interaction features. Furthermore, our approach employs a wavelet kernel to create a set of sub-filters, adeptly denoising by dynamically selecting appropriate filters for the user's evolving characteristics.

% Longitudinally, as the noise in the data increases, the latter two models are significantly less capable of filtering noise than our proposed method.
% In contrast, the random walk-based method is significantly inferior to the spectral GCN method.
% This is mainly because spectral graph neural networks define filters that denoise the useless features in the user-job interactions.
% In addition, our method uses a wavelet kernel to define a series of sub-filters, which is capable of denoising noise by adaptively employing reasonable filters for the dynamically changing features of the user.

%%%%%%%%%%%%%%%%%%%%%%%%%%%%%%%%%%%%%%%%%%%%%%%%%%%
% \noindent \textbf{The validity of hypergraph wavelet filter} \quad
% \subsection{Efficiency Study}
% \noindent \textbf{The training cost} \quad
% \noindent \textbf{Online inference efficiency} \quad
%%%%%%%%%%%%%%%%%%%%%%%%%%%%%%%%%%%%%%%%%%%%%%%%%%%

\subsection{Parametric Study \emph{(RQ4)}}

\noindent \textbf{The size of user (job) groups} \quad
\
The size of user and job groups are two hyperparameters that need to be predefined.
Therefore, we choose 500:1, 1000:1, and 2000:1 as the ratios of the total number of users and the number of user groups $\xi_{u}$, and 100:1, 500:1, 1000:1 as the ratios of the total number of jobs and the number of job groups  $\xi_{v}$ for our experiments respectively, as shown in Figure \ref{fig:para_user} and \ref{fig:para_job}.
We can easily observe that our model achieves best when $\xi_{u} = $1000:1 and $\xi_{v} = $500:1.

\noindent \textbf{The order of Chevbyshev approximation} \quad
\
The order of Chevbyshev approximation greatly impacts the performance of hypergraph wavelet neural networks.
To find the best order, we test our model with $p=1,2,3,4$ and $5$, and the results are shown in Figure \ref{fig:para_p}.
We can see that the performance of the model remains constant when ${p}$ is greater than or equal to 3. Notice that as ${p}$ increases, the computational overhead of the model will also increase, so we choose $p=3$ as the hyperparameter of our model.

\begin{figure}[t]
\centering
\subfigure[Statistics of different session length]{
    \label{fig:session_len_a}
    \resizebox{0.45\linewidth}{!}{
        \begin{tikzpicture}
            \begin{axis}[
                ybar,
                grid=major,
        	xlabel=Session Length,
                ylabel=Frequency ($\times10^5$),
                ymin=0,
                bar width=0.5pt,
                xmin=0,
                xmax=100,
                % nodes near coords,
                % enlarge x limits=0.3,
                % xtick = {1,2,3},
                % xticklabels={500:1,1000:1,2000:1},
                % ylabel=Value,
                % legend columns=2,
                % legend style={font=\large},
                font=\huge
                ]1
                \addplot [fill=mycolor55light, draw=mycolor55, postaction={pattern=crosshatch,pattern color=mycolor55}] table [x=x,y=y,,col sep=comma] {data/exp_session_freq.txt};
                % \addplot [fill=mycolor42light2, draw=mycolor42, postaction={pattern=grid, pattern color=mycolor42}] table [x=x,y=MRR,,col sep=comma] {data/exp_para_user.txt};
                % \legend{H@10, M@10}
            \end{axis}
        \end{tikzpicture}
    }
}
\subfigure[Model performance under different session length]{
\label{fig:session_len_b}
\resizebox{0.475\linewidth}{!}{
\begin{tikzpicture}
\begin{axis}[
	smooth,
	grid=major,
	xlabel=Session Length,
	ylabel=H@10,
	xmin=0,
	xmax=100,
        ymin=0,
        ymax=1,
	legend columns=1,
	legend style={font=\large,at={(0.23,0.7)},anchor=south},
	font=\huge
	]
	\addplot [each nth point=2, mark=*,mycolor55, line width=2pt] table [x=x, y=y1,, col sep=comma] {data/exp_session_hit.txt};
    \addplot [each nth point=2, mark=triangle*,mycolor42, line width=2pt] table [x=x, y=y2,, col sep=comma] {data/exp_session_hit.txt};
    \addplot [each nth point=2, mark=diamond*,mycolor51, line width=2pt] table [x=x, y=y3,, col sep=comma] {data/exp_session_hit.txt};
	\legend{\name, MvDGAE, TiCoSeRec}
\end{axis}
\end{tikzpicture}
}}
\centering
\caption{Results under different session lengths.}
\label{fig:session_len}
\end{figure}
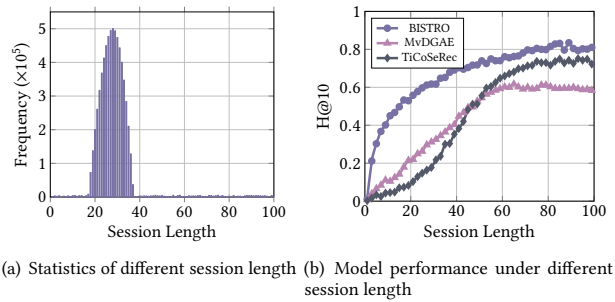

\noindent \textbf{The average length of a session} \quad
\
The length of the session is another hyperparameter that affects the performance of the model.
In Figure \ref{fig:session_len_a}, we can see that the average length of each session is 19-37 on average, and such short behavioral sequences in job recommendations (In job recommendations, the average number of interactions for users to find a suitable job is more than 80 times) are easily interfered with by noisy interactions.
Therefore, we further compared our proposed framework with the top-2 baselines under different session length, as illustrated in Figure \ref{fig:session_len_b}.
It can be seen that when the session length is relatively short, noise has a huge negative impact on the accuracy of all models. However, as the session length decreases, our framework is more robust than the other two methods and can better resist noise interference.

%%%%%%%%%%%%%%%%%%%%%%%%%%%%%%%%%%%%%%%%%%%%%%%%%%%
% \subsection{Case Study}
%%%%%%%%%%%%%%%%%%%%%%%%%%%%%%%%%%%%%%%%%%%%%%%%%%%

\section{Conclusion}
\label{sec:cc}

This study introduces \name, an innovative framework designed to navigate the challenges of job preference drift and the subsequent data noise.
% within the dual-selection dynamics of users and jobs.
The framework is structured around three modules: a coarse-grained semantic clustering module, a fine-grained job preference extraction module, and a personalized top-$k$ job recommendation module. 
Specifically, a hypergraph is constructed to deal with the preference drift issue and a novel hypergraph wavelet learning method is proposed to filter the noise in interactions when extracting job preferences.
The effectiveness and clarity of \name\ are validated through experiments conducted with both offline and online environments.
Looking ahead, we aim to continue refining \name\ to enhance its applicability in broader contexts, particularly in scenarios characterized by anomalous data.

% \begin{acks}
% This research was partially supported by Research Impact Fund (No.R1015-23), APRC - CityU New Research Initiatives (No.9610565, Start-up Grant for New Faculty of CityU), CityU - HKIDS Early Career Research Grant (No.9360163), Hong Kong ITC Innovation and Technology Fund Midstream Research Programme for Universities Project (No.ITS/034/22MS), Hong Kong Environmental and Conservation Fund (No. 88/2022), and SIRG - CityU Strategic Interdisciplinary Research Grant (No.7020046, No.7020074), Huawei (Huawei Innovation Research Program), Tencent (CCF-Tencent Open Fund, Tencent Rhino-Bird Focused Research Program), Ant Group (CCF-Ant Research Fund, Ant Group Research Fund), CCF-BaiChuan-Ebtech Foundation Model Fund, and Kuaishou.
% \end{acks}


\appendix
\balance

\section{Notations}

We summarize all notations in this paper and list them in Table \ref{tab:notation}.

\begin{table}[htb!]
\caption{Notations in this paper.}
\label{tab:notation}
\renewcommand\tabcolsep{5pt}
\begin{filecontents*}{data/notation.txt}
Notation; Description
$\mathcal{S}^u$;The set of session-based interactions
$\mathcal{G}$;Interaction graph
$\mathcal{V}$;Group-based node-set, $\mathcal{V} = \{ v_1, v_2, \cdots \}$
$\mathcal{E}$;The hyperedge set, $\mathcal{E} = \{ e_1, e_2, \cdots \}$
$\mathcal{U}$;The user set, $\mathcal{U} = \{ u_1, u_2, \cdots \}$
$\mathcal{X}$; Graph features that extracted from user/job contents
$\boldsymbol{D}^u$; The user resume set
$\boldsymbol{D}^v$; The job requirement set
$\boldsymbol{C}_u \& \boldsymbol{C}_v$; User/Job groups
$\mathcal{L}$;Graph Laplacian matrix
$\boldsymbol{H}$;The node-edge relationship matrix of the hypergraph
$\boldsymbol{D}_v \& \boldsymbol{D}_e$;The degree matrix of nodes and hyperedges
$\boldsymbol{U}$;Eigen-vectors of graph Laplacian matrix
$\boldsymbol{g}_\theta$;Graph filter with training parameter $\theta$
$k$, The number of jobs generated by the job recommender system at once
$\xi$; The ratio of the users (jobs) to user (job) groups
$p$; The order of Chebyshev approximation
\end{filecontents*}
\csvreader[
separator=semicolon,
  tabular=cc,
  table head=\hline
 \multicolumn{1}{c}{\bfseries{Notation}} & \multicolumn{1}{c}{\bfseries{Description}} \\
  \hline,
  late after last line= \\
  \hline % horizontal line at the end of the table
]{data/notation.txt}{}{\csvlinetotablerow}
\end{table}

\section{Model Complexity}

Our proposed framework, BISTRO, is efficient. We will analyze it from two aspects: theoretical [from $O(n^3)$ to $O(n\log n)$] and application (less than $100ms$ per sample) level.

\noindent \textbf{Theoretical level}

Among all modules in BISTRO, the graph neural network (GNN) is considered very time-consuming. Actually, graph learning-based recommender systems have computational limitations in practice. To address this issue, we cluster users (jobs) based on the semantic information in their resumes (requirements) using the K-Means algorithm and use a simple RNN to extract the personalized preference for a person in the user group. The extraction of preference features based on user (job) groups reduces the computational overhead of GNNs. Noting that eigendecomposition in GNNs is resource-intensive, we place it by using the Chebyshev polynomial estimation, and the Chebyshev coefficient in the polynomial can be computed by Fast Fourier Transform, which reduces the computational complexity from exponential complexity $O(n^3)$ to $O(n\log n)$. Therefore, our algorithm is efficient.

\noindent \textbf{Application level}

In practice, the training of all models is performed offline. For example, we use spark cluster to calculate the clustering center of each group and use HGNN to learn the corresponding representation of groups. For new or updated users and jobs, we assign them to the nearest group based on semantic clustering. Only the RNN module operates online, inferring personalized user representations within groups. In the online experiment, the 99\% Response Time of BISTRO is less than $100ms$.

\section{Experiment Detail}
\label{app:sim}

\subsection{Baselines Detail}
\label{app:baselines}

The details of these baselines are as follows:

\noindent $\bullet$ BasicMF \cite{koren2009matrix}: A model that combines matrix factorization with a Multilayer Perceptron (MLP) for recommendations.

\noindent $\bullet$ ItemKNN \cite{wang2006unifying}: A recommender that utilizes item-based collaborative filtering.

\noindent $\bullet$ PureSVD \cite{cremonesi2010performance}: An approach that applies Singular Value Decomposition for recommendation tasks.

\noindent $\bullet$ SLIM \cite{ning2011slim}: A recommendation method known as the Sparse Linear Method.

\noindent $\bullet$ DAE \cite{wu2016collaborative}: Stands for Collaborative Denoising Auto-Encoder, used in recommendation systems.

\noindent $\bullet$ MultVAE \cite{liang2018variational}: A model extending Variational Autoencoders to collaborative filtering for implicit feedback.

\noindent $\bullet$ EASE \cite{steck2019embarrassingly}: A recommendation technique called Embarrassingly Shallow Autoencoders for Sparse Data.

\noindent $\bullet$ P3a \cite{cooper2014random}: A method that uses ordering rules from random walks on a user-item graph.

\noindent $\bullet$ RP3b \cite{paudel2016updatable}: A recommender that re-ranks items based on 3-hop random walk transition probabilities.

\noindent $\bullet$ NGCF \cite{wang2019neural}: Employs graph embedding propagation layers to generate user/item representations.

\noindent $\bullet$ LightGCN \cite{he2020lightgcn}: Utilizes neighborhood information in the user-item interaction graph.

\noindent $\bullet$ SLRec \cite{yao2021self}: A method using contrastive learning among node features.

\noindent $\bullet$ SGL \cite{wu2021self}: Enhances LightGCN with self-supervised contrastive learning.

\noindent $\bullet$ GCCF \cite{chen2020revisiting}: A multi-layer graph convolutional network for recommendation.

\noindent $\bullet$ NCL \cite{lin2022improving}: Enhances recommendation models with neighborhood-enriched contrastive learning.

\noindent $\bullet$ DirectAU \cite{wang2022towards}: Focuses on the quality of representation based on alignment and uniformity.

\noindent $\bullet$ HG-GNN \cite{pang2022heterogeneous}: Constructs a heterogeneous graph with both user nodes and item nodes and uses a graph neural network to learn the embedding of nodes as a potential representation of users or items.

\noindent $\bullet$ A-PGNN \cite{zhang2020personalized}: Uses GNN to extract session representations for intra-session interactions and uses an attention mechanism to learn features between sessions.

\noindent $\bullet$ AdaGCL \cite{jiang2023adaptive}: Combines a graph generator and a graph denoising model for contrastive views.

\noindent $\bullet$ MvDGAE \cite{zheng2021multi}: Stands for Multi-view Denoising Graph AutoEncoders.

\noindent $\bullet$ STAMP \cite{DBLP:conf/kdd/LiuZMZ18}: A model based on the attention mechanism to model user behavior sequence data.

\noindent $\bullet$ GRU4Rec \cite{DBLP:journals/corr/HidasiKBT15}: Utilizes Gated Recurrent Units for session-based recommendations.

\noindent $\bullet$ BERT4Rec \cite{DBLP:conf/cikm/SunLWPLOJ19}: A model for the sequence-based recommendation that handles long user behavior sequences.

\noindent $\bullet$ CL4Rec \cite{xie2022contrastive}: An improved version of BERT4Rec with locality-sensitive hashing for faster item retrieval.

\noindent $\bullet$ CoScRec \cite{DBLP:journals/corr/abs-2108-06479}: It explores an innovative recommendation approach that enhances sequential recommendation systems through robust data augmentation and contrastive self-supervised learning techniques.

\noindent $\bullet$ TiCoSeRec \cite{DBLP:conf/aaai/DangYGJ0XSL23}:  A method based on CoSeRec, utilizing data augmentation algorithms for sequence recommendation improvement.

\subsection{The number of recommended jobs}
\label{app:para_k}

The hyperparameter, $k$, also has a critical impact on experimental results.
We set $k=5,10$ and $20$ to conduct experiments respectively.
The experimental results are shown in Table \ref{tab:para_k}.
It can be seen our model performs well under all settings.

\begin{table}[t]
 \caption{Results under different settings of $k$.}
 \label{tab:para_k}
 \begin{filecontents*}{data/exp_para_k.txt}
title,col1,col2,col3
\textbf{H@$k$},0.6059,0.6247,0.6303
\textbf{M@$k$},0.5117,0.5131,0.5051
\end{filecontents*}
\csvreader[
   tabular=cccc,
   table head=\hline
  
   \specialrule{0em}{1pt}{1pt}

   \multicolumn{1}{c}{} & \multicolumn{1}{c}{\bfseries{$k=5$}} & \multicolumn{1}{c}{\bfseries{$k=10$}} & \multicolumn{1}{c}{\bfseries{$k=20$}}
  
   \\ \specialrule{0em}{1pt}{1pt}
  
   \hline,
   late after line={\\},
   late after last line= \\
   \hline % horizontal line at the end of the table
 ]{data/exp_para_k.txt}{}{\csvlinetotablerow}
 \end{table}

\subsection{Case Study}
\label{app:case}

\begin{figure}[h]
\tikzstyle{user} = [circle, draw=mycolor44, fill=mycolor44light, drop shadow, align=center]
\tikzstyle{job} = [diamond, draw=mycolor43, fill=mycolor43light, drop shadow, align=center]
% , fill opacity=0.5
\tikzstyle{contentL}=[rectangle, rounded corners, minimum width=6em, minimum height=6em, text centered, draw=mycolor55, fill=mycolor55light, drop shadow, align=left, text=white]
\tikzstyle{contentM}=[rectangle, rounded corners, minimum width=6em, minimum height=4em, text centered, draw=mycolor55, fill=mycolor55light, drop shadow, align=left, text=white]
\tikzstyle{contentS}=[rectangle, rounded corners, minimum width=1em, minimum height=1em, text centered, draw=mycolor42, fill=mycolor42light, drop shadow, align=center, text=white]

\resizebox{\linewidth}{!}{
\begin{tikzpicture}[thick, auto, node distance=1.3cm,>=latex']

\path

(0,0.7) node [contentL] (c3) {\ \\ \ \\Skill Req: bb+cc\\Degree Req: dd\\Experience: hh}
(0,1.3) node [contentS] (c5) {Location: gg}
(2.4,0.7) node [job] (j3) {JID\\***872}
(4.8,0.7) node [job] (j4) {JID\\***994}
(7.2,0.7) node [contentL] (c4) {\ \\ \ \\Skill Req: ee+ff\\Degree Req: dd\\Experience: hh}
(7.2,1.3) node [contentS] (c6) {Location: gg}

(0,3.5) node [job] (j1) {JID\\***265}
(2.4,3.5) node [user] (u10) {UID\\***175}
(4.8,3.5) node [user] (u11) {UID\\***175}
(7.2,3.5) node [job] (j2) {JID\\***523}

(0,6) node [user] (u2) {UID\\***479}
(2.4,6) node [contentM] (c1) {Age: aa\\Skill: bb, cc\\Degree: dd}
(4.8,6) node [contentM] (c2) {Age: aa\\Skill: ee, ff\\Degree: dd}
(7.2,6) node [user] (u3) {UID\\***013}

(3.6,0) node [align=center] (newj) {New Jobs}
(3.6,3.5) node [align=center] (text1) {Revise\\Resume}
(1.4,2.2) node (exp1) {\rotatebox{45}{Expect}}
(5.8,2.2) node (exp2) {\rotatebox{-45}{Expect}}
(1.4,3.7) node (click1) {Click}
(6,3.7) node (rec3) {Rec.}
(0.2,4.9) node (click3) {\rotatebox{90}{Click}}
(7,4.9) node (click4) {\rotatebox{-90}{Click}}
(2.6,2.2) node (rec1) {\rotatebox{90}{Rec.}}
(4.6,2.2) node (rec2) {\rotatebox{-90}{Rec.}}
;

\draw[->,black,thick] (u10) -- (j1);
\draw[->,black,thick] (u10) -- (u11);
\draw[->,black,thick] (u11) -- (c2);
\draw[->,black,thick] (j1) -- (c3);
\draw[->,black,thick] (u2) -- (j1);
\draw[->,black,thick] (u2) -- (c1);
\draw[->,black,thick] (u10) -- (c1);
\draw[->,black,thick] (u3) -- (c2);
\draw[->,black,thick] (j2) -- (u11);
\draw[->,black,thick] (j2) -- (c4);
\draw[->,black,thick] (j4) -- (c4);
\draw[->,black,thick] (u11) -- (c6);
\draw[->,black,thick] (u10) -- (c5);
\draw[->,black,thick] (j4) -- (u11);
\draw[->,black,thick] (j3) -- (u10);
\draw[->,black,thick] (j3) -- (c3);
\draw[->,black,thick] (u3) -- (j2);

\end{tikzpicture}}

\caption{A real-life scenario for the job recommendation.}
\label{fig:cs}
\end{figure}
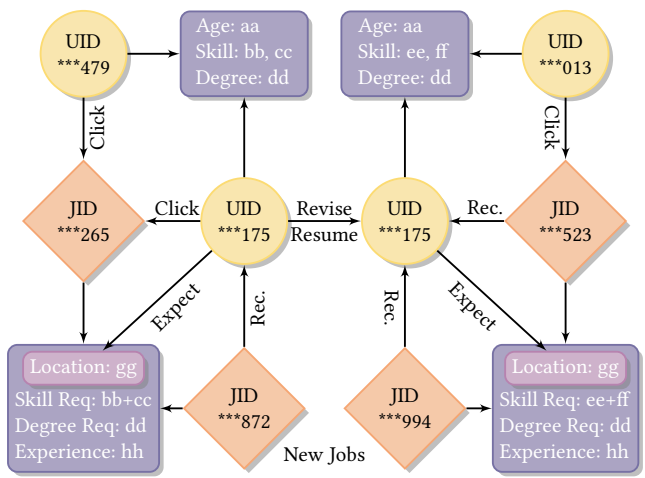

Beyond its effectiveness in performance, \name\ also boasts considerable interpretability.
To demonstrate how the framework mitigates both the job preference drift and data noise problems, we present a real-life scenario to illustrate the logic behind the suggestions made by \name, as shown in Figure \ref{fig:cs}.

In this figure, jobs with IDs ***872 and ***994 are two job positions that are newly posted in the online recruitment system, while IDs ***265 and ***523 are two job positions that a large number of users interact with frequently.
Among them, ***872 and ***265, as well as ***994 and ***523, have similar occupational demand descriptions respectively.
Also, the user with ID ***175 shared a similar resume with user ID ***479 before ***175 modified the resume, and after his resume was changed, ***175 had a similar content with user ***013.
Recommendations in this scenario can be divided into three examples:

\noindent\textbf{Example 1} (Recommendation for a dynamically changing user)\quad
\
Consider the user represented by ID ***175, \name\ addresses this challenge by deploying content-based analysis.
The framework utilizes the user's social network and a set of resume attributes collected to create a composite feature profile to identify users with similar tastes.
Subsequently, it recommends a job with ID ***523 favored by a like-minded user with ID ***013 to him.

\noindent\textbf{Example 2} (Recommendation for a new job)\quad
\
A newly posted job with ID ***872 lacks any user interaction data, complicating the generation of a meaningful representation for it. 
\name, however, overcomes this by incorporating auxiliary information such as skill requirements and working experience, and then associated tags to locate similar content.
By leveraging this approach combined with the user's expectations, \name\ acquires a rich and informative embedding for the job, enabling it to recommend the job to users who have shown an interest in comparable jobs.

\noindent\textbf{Example 3} (Recommend a new job to a dynamically changing user)\quad
\
Combining both two situations illustrated above, \name\ deals with this complex challenge by utilizing a wavelet graph denoising filter and graph representation method.
In this way, it can recommend the latest jobs with similar job content to users with the same real-time needs as well as similar user content characteristics.

\begin{table}[!ht]
\caption{Results of different job recommender systems.}
\label{tab:app_case}
\renewcommand\tabcolsep{2.5pt}
\begin{filecontents*}{data/app_case.txt}
title,col1,col2,col3,col4,col5,col6
InEXIT,0.4131,0.2936,0.4611,0.3762,0.5141,0.4033
DGMN,0.4274,0.3178,0.4897,0.3992,0.5217,0.4125
APJFMF,0.4352,0.3114,0.4863,0.3910,0.5254,0.4166
\textbf{Ours},\textbf{0.5598},\textbf{0.5467},\textbf{0.6166},\textbf{0.5152},\textbf{0.6247},\textbf{0.5131}
\end{filecontents*}
\csvreader[
  tabular=ccccccc,
  table head=\hline
  
  \specialrule{0em}{1pt}{1pt}

  \multicolumn{1}{c}{} & \multicolumn{2}{c}{\bfseries{Shenzhen}} & \multicolumn{2}{c}{\bfseries{Shanghai}} & \multicolumn{2}{c}{\bfseries{Beijing}}
  
  \\ \specialrule{0em}{1pt}{1pt}

  \multicolumn{1}{c}{} & \multicolumn{1}{c}{\bfseries{H@10}} & \multicolumn{1}{c}{\bfseries{M@10}} & \multicolumn{1}{c}{\bfseries{H@10}} & \multicolumn{1}{c}{\bfseries{M@10}} & \multicolumn{1}{c}{\bfseries{H@10}} & \multicolumn{1}{c}{\bfseries{M@10}}
  
  \\ \specialrule{0em}{1pt}{1pt}
  
  \hline,
  late after line={\\},
  late after last line= \\
  \hline % horizontal line at the end of the table
]{data/app_case.txt}{}{

\csvlinetotablerow
}

\small{
Bold indicates the statistically significant improvements \\ (\ie two-sided t-test with p < 0.05) over the best baseline (underlined).

For all metrics: the higher, the better.
}

\end{table}

In addition, we also compare the proposed framework, \name, with multiple job state-of-the-art recommender systems, \ie InEXIT \cite{shao2023exploring}, DGMN \cite{bian2019domain}, and APJFMF \cite{jian2024your}.
The result can be found in Table \ref{tab:app_case}.
We can see that our framework acheves the best among all baselines, which verify the effectiveness of our method.

\end{document}